\documentclass[]{aa}
\usepackage{graphicx}
\usepackage{natbib}  
\bibpunct{(}{)}{;}{a}{}{,}
\usepackage{txfonts}
\usepackage{hyperref}

\defcitealias{Lazorenko2009}{PL09}
\defcitealias{JS2013}{Paper\,I}

\newcommand{\dwone}{DE0615$-$01}

\newcommand{\dwfive}{DE0716$-$06}

\newcommand{\dwnine}{DE0823$-$49}

\newcommand{\dwfourt}{DE1253$-$57}

\newcommand{\dwtwenty}{DE1756$-$48}

\begin{document}

 \title{Astrometric planet search around southern ultracool dwarfs}
   \subtitle{II. Astrometric reduction methods and a deep astrometric catalogue\thanks{\textit{Based on observations made with ESO telescopes at the La Silla Paranal Observatory under programme IDs 086.C-0680, 087.C-0567, 088.C-0679, 089.C-0397, and 090.C-0786.}}}

\author{P. F. Lazorenko\inst{1}
		\and J. Sahlmann\inst{2,3}
		\and D. S\'egransan\inst{3}
		\and E. L. Mart\'in\inst{4} 
	        \and M. Mayor\inst{3} 
		\and D. Queloz\inst{3,5} 
		\and S. Udry\inst{3}}	

\institute{Main Astronomical Observatory, National Academy of Sciences 
of the Ukraine, Zabolotnogo 27, 03680 Kyiv, Ukraine\\
		\email{laz@mao.kiev.ua}				
		\and
		European Space Agency, European Space Astronomy Centre, P.O. Box 78, Villanueva de la Ca\~nada, 28691 Madrid, Spain
		\and
		Observatoire de Gen\`eve, Universit\'e de Gen\`eve, 51 Chemin Des Maillettes, 1290 Versoix, Switzerland
		\and
		INTA-CSIC Centro de Astrobiolog\'ia, 28850 Torrej\'on de Ardoz, Madrid, Spain
		\and
		University of Cambridge, Cavendish Laboratory, J J Thomson Avenue, Cambridge, CB3 0HE, UK
}

\date{Received 18 December 2013 / Accepted 14 March 2014}
\abstract{}
{We describe the astrometric reduction of images obtained with the FORS2/VLT camera in the framework of an astrometric planet search around 20 M/L-transition dwarfs. We present the correction of systematic errors, the achieved astrometric performance, and a new astrometric catalogue containing the faint reference stars in 20 fields located close to the Galactic plane.}
{Remote reference stars were used both to determine the astrometric trajectories of the nearby planet search targets and to identify and correct systematic errors.}
{We detected three types of systematic errors in the FORS2 astrometry: the relative motion of the camera's two CCD chips, errors that are correlated in space, and an error contribution of as yet unexplained origin. The relative CCD motion probably has a thermal origin and typically is  0.001--0.010~px ($\sim$0.1--1~mas), but sometimes amounts to 0.02--0.05~px (3--6 mas). This instability and space-correlated errors are detected and mitigated using reference stars. The third component of unknown origin has an amplitude of 0.03--0.14~mas and is independent of the observing conditions.
We find that a consecutive sequence of 32 images of a well-exposed star over 40 min at 0.6\arcsec~seeing results in a median r.m.s. of the epoch residuals of 0.126~mas. Overall, the epoch residuals are distributed according to a normal law with a $\chi^2$ value near unity.
We compiled a catalogue of 12\ 000 stars with $I$-band magnitudes  of 16--22 located in 20 fields, each covering $\sim 2\arcmin\times2\arcmin$. It contains  $I$-band magnitudes, ICRF positions with 40--70 mas precision,  and relative proper motions and absolute trigonometric parallaxes with a precision of 0.1~mas/yr and 0.1~mas at the bright end, respectively.}
{This work shows that an astrometric accuracy of $\sim$100 micro-arcseconds over two years can be achieved with a large optical telescope in a survey covering several targets and varying observing conditions}

\maketitle

\section{Introduction}{\label{notes}}
Extrasolar planets around stars can be discovered and characterised by a variety of observation techniques \citep{Seager:2011ve}. Different methods give access to different observables of a given exoplanetary system and are subject to different practical limitations. To obtain a complete picture of extrasolar planets, we therefore depend on having access to the widest possible range of observing techniques.

Astrometry consists in measuring the photocentre positions of stellar objects and can be used for indirect exoplanet detection by revealing a star's orbital reflex motion \citep{Sozzetti:2005qy}. However, this relies on a measurement accuracy better than one milli-arcsecond (mas) over time-scales of several years, which requires specialised instruments and methods. Dedicated space missions such as \emph{Hipparcos} \citep{ESA:1997vn} and, in particular, its successor \emph{Gaia} \citep{Perryman:2001vn, de-Bruijne:2012kx} can meet these requirements, and instruments on-board \emph{Hubble} have been used for this purpose, too \citep{Benedict:2010ph}.
On the ground, the main practical limitation is turbulence in the Earth's atmosphere \citep{Sahlmann_spie}, which can be overcome with the use of large-aperture telescopes, interferometry, or adaptive optics.

For observations of dense stellar fields with the optical camera {\small FORS2} of ESO's Very Large Telescope, we have achieved astrometric precisions of 0.05~mas for well-exposed star images at the field centre when the seeing is restricted to the optimal range { (\citealt{Lazorenko2009},  hereafter \citetalias{Lazorenko2009})}. Because this performance is sufficient for exoplanet detection, we initiated an astrometric search targeting ultracool dwarfs { (UCD)} at the M/L transition in 2010, a survey that is described in detail in the first paper in our series { (\citealt{Sahlmann:2013prep}, hereafter \citetalias{JS2013}).}

The detection of planetary signals at or close to the noise level requires high-quality astrometric data, that is, precise position measurements ideally free of systematic errors, and adequate precision estimates that correspond to the measurement accuracy. In this paper, we present the latest optimisations of the astrometric methods for the reduction of {\small FORS2} observations described in { \citetalias{Lazorenko2009}.  These methods were already successfully used for the detection of a close 28 Jupiter-mass ($M_J$) companion of an L1.5 dwarf \citep{JS2013} and are implemented for the planet search survey \citepalias{JS2013}. }

The paper is structured as follows: the observations are described in Sect. \ref{obs}, the astrometric model is outlined in Sect. \ref{mod}, and the single-frame precision of the measurements is characterised in Sect. \ref{fr_pr}. The epoch residuals and  precision of these residuals are introduced in Sect. \ref{ep_res}, and in Sect. \ref{se} we demonstrate the method of the detection and reduction of systematic errors. The observed $\chi^2$ statistic for the epoch residuals and its comparison with the theoretical $\chi^2$ distribution law is analysed in Sect. \ref{conc}.  In Sect. \ref{cat}, we describe the catalogue of astrometric and photometric data obtained for field stars in the observed fields. We conclude in Sect. \ref{c_d}.

\section{Observations and initial data reduction}{\label{obs}}
Observations were obtained at ESO's Very Large Telescope (VLT) with the $4.2\arcmin \times 4.2$\arcmin-field imaging camera {\small FORS2} \citep{FORS},  { whose detector is composed of two CCD chips and } has a high-resolution collimator, yielding an image scale of 0.126\,\arcsec\,px$^{-1}$ with $2\times 2$~pixel binning. The target list \citepalias{JS2013} contains 20 southern M/L dwarfs, which were monitored for two years starting in November 2010, each with approximately  4--6 epochs per year with one-month spacing in the visibility period.  In the third year, a few more follow-up observations were obtained for some targets, for example the binary \dwnine~\citep{JS2013}. Because the targets are red dwarfs, we observed with the $I$-Bessel filter with a central wavelength at {768~nm.} We obtained { 9--16} epochs, each consisting of a series of 20 to 60 exposures in $I$ band, for every target. The target list with short identifiers (ID) from \citetalias{JS2013}, which refer to approximate field sky position, the $I$-band magnitudes (see Sect. \ref{ph}), and the number of observing epochs $N_e$ are given in Table \ref{ident}. A more detailed description of the observations is given in \citetalias{JS2013}.

\begin{table}[tbh]
\caption [] {Target list.}
\centering
\begin{tabular}{@{}cccc|cccc@{}}
\hline
\hline
Nr&     ID &  $m_I$  & $N_e$&  Nr &     ID   &$m_I$  & $N_e$ \rule{0pt}{11pt}\\
\hline
1 & DE0615-01& 17.0  &  11  & 11  & DE1048-52& 17.5  &  12  \rule{0pt}{11pt} \\
2 & DE0630-18& 15.7  &  16  & 12  & DE1157-48& 17.3  &  10 \\
3 & DE0644-28& 16.9  &  11  & 13  & DE1159-52& 14.6  &   9 \\
4 & DE0652-25& 16.0  &  11  & 14  & DE1253-57& 16.7  &  10 \\
5 & DE0716-06& 17.5  &  10  & 15  & DE1520-44& 16.8  &  10 \\
6 & DE0751-25& 16.5  &  12  & 16  & DE1705-54& 16.5  &  11 \\
7 & DE0805-31& 16.0  &  11  & 17  & DE1733-16& 16.9  &  10 \\
8 & DE0812-24& 17.2  &  11  & 18  & DE1745-16& 17.0  &  10 \\
9 & DE0823-49& 17.1  &  15  & 19  & DE1756-45& 15.5  &  10 \\
10& DE0828-13& 16.1  &  11  & 20  & DE1756-48& 16.7  &  10 \\
\hline
\end{tabular}
\label{ident}
\end{table}

The ESO archive contains 7970 exposures of our programme collected during five semesters. Of these, we effectively used 6813 exposures, meaning that there is a nominal overhead of $\sim$14\%. This is mostly attributable to observations taken in poor conditions that were repeated, but nevertheless appear in the archive. In general, they did not pass the reduction process, which rejects frames with image FWHM larger than 0.9\arcsec. The overhead also includes the 4\,\% of exposures for each target that were rejected to remove  outliers and the data with lowest precision.

The raw images were de-biased and flat-fielded using the standard calibration files. The computation of photocentres is based on fitting the stellar image profiles with a model composed of three centred  Gaussian components. Due to the complex shape of the measured point spread function (PSF), deviations from the analytic model often exceed the fluctuations expected from the Poisson statistics of photoelectrons and contain a systematic component common to all star images within a frame. A model of these deviations was compiled for each exposure and subtracted from the measured images of stars, ensuring a significantly better fit and a higher precision of the photocentre determination \citep{Lazorenko2006}.

This procedure was improved for crowded fields, where close pairs of star images were processed iteratively with mutual subtraction of light coming from the nearby image component. This required modelling  the PSF variation over each of the two CCD chips, which were processed separately. For this purpose, we divided the detector field into a grid of 16 $\times$ 16 square areas. In each area, we selected a few dozen isolated bright stars to create two-dimensional PSFs by co-adding these images with a resolution of 1/2~px and assigning a weight to each star that decreased with increasing distance from the grid node. In this way, we obtained a set of average PSFs that are evenly spaced over the CCD. For each CCD chip, a polynomial fit was used to derive the final calibration PSF with a size of $30\times 30$~px ($3.8\times 3.8$\arcsec) at 1/2~px resolution and a model of its deformation across the CCD. The PSF estimate is good up to 15~px from the star centre, but becomes too noisy at larger distances, where we applied a simple axially symmetric approximation up to 40~px. This calibration of the PSF allowed us to process crowded fields with stellar densities of 300 stars/arcmin$^2$.

\subsection{Saturated target images}{\label{satur}}
At the start of the survey and in very good seeing conditions ($\lesssim$0.5\arcsec), it occurred that the central pixels of bright star images were saturated, which increased the photocentre uncertainties and, even worse, introduced systematic displacements. Only a few percent of the exposures were affected, but it produced strongly biased solutions when the processing was made with a simple rejection of saturated pixels. For instance, with four saturated pixels, the bias in the photocentre position was about 1~mas, significantly higher than the epoch precision of 0.1~mas. To remove this effect, we modelled it by artificially saturating field-star images with the same saturation pattern as for the target. The difference in photocentre position between the original and modified images, averaged over all field stars, yields the bias value for the target. This method, applied to the target alone, reduced the systematic epoch-averaged image displacements to 0.1~mas for one saturated pixel, which is similar to the epoch precision. However, for four saturated pixels, the residual bias of 0.2--0.3~mas was just at the tolerance limit. Therefore, images with five or more saturated pixels in the target image were rejected. Saturated images of reference stars were always rejected to exclude possible systematic errors, at the expense of a slight increase in the reference frame noise.

After  recognising these problems caused by saturation, we shortened the exposure $T$ durations in the survey observations to prevent saturation in very good seeing\footnote{In this paper, \emph{seeing} refers to the image FWHM, which are usually considered to be different quantities \citep{seeing}.}. We selected exposure durations that resulted in $0.5-0.7 \times 10^6$ photoelectrons per target image, which corresponds to the light collected in a fully exposed image at best seeing. For average seeing of 0.6\arcsec, this results in only one third of the maximum flux, but it ensures a low probability of saturation, thus preventing the appearance of systematic errors.

\section{Astrometric model}
\begin{figure}[h!]
\resizebox{\hsize}{!}{\includegraphics*[bb = 113 247 500 547, width=\linewidth]{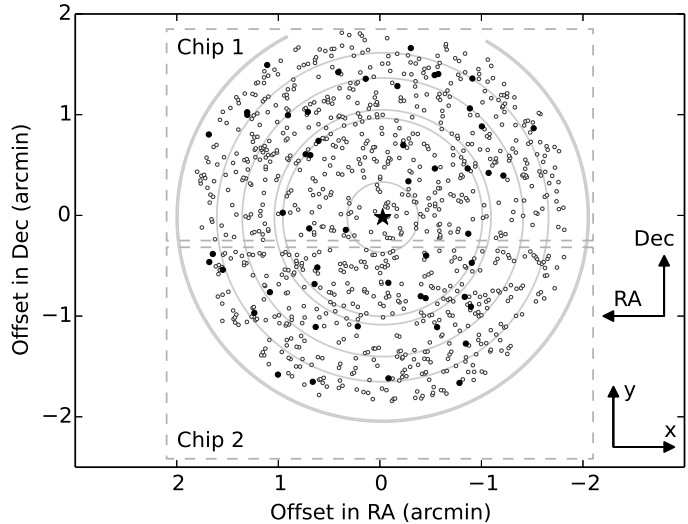}}
\caption{Stars in the \dwtwenty\ field that are used for the astrometric reduction and are included in the catalogue. The UCD is marked with a star, bright stars with magnitudes within $\pm$0.5 mag of the target are shown with solid circles, and open circles show all other reference stars. The footprints of the FORS2 chips and the axes orientations are indicated. Only { catalogue stars within 
1\arcmin 50\arcsec around the target are shown, 
and concentric circles with increasing radii of $R_{\rm opt} (k)$ mark reference fields 
used  for for the reduction with $k=4,6,8,10,12,14,16$. See text for details.}
}
\label{refStars}
\end{figure}

{\label{mod}}
The   astrometric reduction is based on the symmetrising method applied to reduce atmospheric image motion {  (\citet{LazLaz}; \citetalias{Lazorenko2009}).  The implementation of the method, however, has significantly changed and several crucial improvements were made.} Figure \ref{refStars} shows a typical configuration of stars in the focal plane of {\small FORS2} and illustrates some of the concepts and conventions discussed in the next sections.

\subsection{Atmospheric image motion}{\label{im_mot}}
The atmospheric image motion has a continuous spectrum of spatial frequencies, that can be represented by an infinite expansion in series of even integer powers $k=2,4,...$, where $k$ is the modal index, and the mode amplitude decreases rapidly with increasing $k$. Over the CCD space ($x,y$),  image motion causes deformations of the frame of reference stars  that for each frame exposure $m$
are modelled by the expansion  $\Phi_m(x,y)=\sum_{w}c_{w,m}f_{w}(x,y)$
 into a series of basic functions $f_{w}(x,y)$,  whose number $w= 1 \ldots k(k+2)/8$ depends on $k$.  The functions $f_{w}(x,y)$ are full polynomials in $x, y$ of all powers from zero to the highest power $k^*= k/2-1$, and $c_{w,m}$ are the expansion coefficients. 
The centre of { the reference frame} 
is aligned with a star called the \emph {target}, therefore the solution is only valid for this star. 

In practice, we realised a sequence of independent solutions with all modes $k=4,6 \ldots 16$,  which corresponds to the power $k^*$ taking values between unity (a linear  model) and $k^*=7$.
For each solution, the field of reference stars is restricted to the optimal radius $R=R_{\rm opt}(k)$ (e.g.\ Fig.~\ref{refStars}), whose value depends on $k$ and corresponds to the field size where the image motion variance $\sigma_{\rm atm}^2 = B R^b T^{-1}$ 
(which increases with $R$) equals the reference frame noise $\sigma_{\rm rf}^2$ (which decreases with $R$)
\begin{equation}
\label{eq:balance}
\sigma_{\rm atm}^2 = \sigma_{\rm rf}^2.
\end{equation}
The values of the coefficients $B$ and $b$ for each $k$ are given in \citet{Lazorenko2006} and \citetalias{Lazorenko2009}, but they are approximate and allow us  to derive only the first approximation of  $R_{\rm opt}$ by numerical solution of Eq.~(\ref{eq:balance}). Its determination can be improved as  detailed in Sect.~\ref{Ropt}. For targets near the field centre, $R_{\rm opt}$ is typically between 1\arcmin\ and 2\arcmin, when the intermediate mode $k=10$ is applied, whereas for the highest mode $k=16$, the reference frame covers most of the FoV (Fig.~\ref{refStars}). Solutions obtained with different $k$ are highly correlated and are eventually merged to obtain the final solution.

\subsection{Differential chromatic refraction}{\label{dcr}}

Positional observations through the atmosphere are affected by the refraction that deflects the light rays to zenith depending on the effective wavelength and zenith distance $\zeta$. The measured effect is known as differential chromatic refraction (DCR), because it shifts images of blue and red stars differently
 by $\rho \tan \zeta$, where $\rho$ is the difference of refraction indexes of the target and reference stars. The value of $\rho$ can be computed if the spectral energy distribution is known \citep{Monet1992,Pravdo1996}.

 {\small FORS2} observations are obtained with an atmospheric dispersion compensator LADC \citep{Avila}, which produces inverse deflection of light rays  on $d \tan \zeta_{\rm LADC}$, where $d$ is a constant dependent on the star color and $\zeta_{\rm LADC}$ is the effective zenith angle of the LADC. In the current programme, we derive $\rho$ and $d$ as  free model parameters for each star (\citet{Pravdo2004}; \citetalias{Lazorenko2009}).

\subsection{Computational layout of the model}{\label{layout}}

The reduction is run in a sequence of repeating steps, starting from the computation of the positional residuals  (Sect.~\ref{field_d}, \ref{astr_mod_T}) for a single UCD or a set of field stars. This is followed by the computation of  $R_{\rm opt}$, which implies multiple repetitions of the preceding steps (Sect.\ref{Ropt}). The complete sequence is repeated again in the course of improving determination of the systematic corrections (Sect.\ref{res}), because the value of $R_{\rm opt}$ changes.

\subsubsection{Deriving field deformations $\Phi (x,y)$}{\label{field_d}}
Let $X_{i,m}$ be the measured  { two-dimensional CCD positions of the reference star $i$ in the frame (=exposure) $m=1 \ldots M$,  which we assume to include the measurements in both  $x$ and $y$ unless the other (one-dimensional case)} is evident from the context.  The astrometric reduction of these data is performed within a field of the radius $R_{\rm opt}$ and reveals the { deformations $\Phi_m(x,y)$  of the field geometry in frame $m$. The motion of   each of $i= 1 \ldots N_{\rm rf}$ reference star } in time  is described  by  the model motion $\sum_{s=1}^S \xi_{i,s}\nu_{s,m}$ with $S$  astrometric parameters $\xi_{i,s}$ and  known factors $\nu_{s,m}$, for instance the parallax factor, time, etc. The set of parameters $\xi_{i,s}$ contains two zero points, two proper motions,  the parallax, DCR parameter $\rho$, and the LADC parameter $d$, thus, $S=7$ and $s=1..S$. {  The solution for   $S \times N_{\rm rf}$ astrometric parameters  $\xi_{i,s}$  and $2\times M \times k(k+2)/8$ coefficients $c_{w,m}$ 
is derived by the least-squares fit of $2\times M \times N_{\rm rf}$ 
displacements  $\Delta X_{i,m} = X_{i,m}-X_{i,0}$ of reference stars relative to 
the grid of stars within $ R<R_{\rm opt}$  in a reference image $m=0$. The reference star sample and solution of the model is unique for each target. [text deleted]}

{ It follows from the procedures introduced in  Sect.~\ref{se} that the measurements $X$ made in the frames $m$ and $m'$ of the same epoch $e$ (i.e.\ taken in the same night) are correlated. Therefore,
the covariance matrix $ \vec{D}$  of $X$ (or $ \Delta X $) has a block structure with the block-diagonal elements $D_{m,m}=\sigma_m^2$,  where  $\sigma_m$ is the precision of a single frame measurement},
non-zero off-diagonal elements  $ D_{m,m'}$ if $m,m' \in e$, and $ D_{m,m'}=0$ if $m$ and $m'$ were taken in different epochs (Appendix~\ref{B3}).

{ The astrometric parameters $\xi_{i,s}$
are relative because
the solution }
only converges if the $k(k+2)/8$ additional restrictions $\sum_i \xi_{i,s} P_i f_{w}(x_i,y_i)=0$ are set, where $P_i$ are optional weighting factors. In particular, we require
\begin{equation}
\label{eq:z}
 \sum_i P_i\, \mu_{\alpha,i}=\sum_i P_i\, \mu_{\delta,i}=\sum_i P_i \,{\varpi_i}=\sum_i P_i \,{\rho_i}= \sum_i P_i \,{d_i}=0,
\end{equation}
which implies that approximately half of the relative parallaxes $\varpi$ of reference stars are negative and that the average of all proper motions is zero. 

\subsubsection{Deriving astrometric parameters for the target star}{\label{astr_mod_T}}
{ Using the solution of the astrometric model,
we computed $2\times M$ displacements $\Delta X_{m} = X_{m}-X_{0}- \Phi_{m}(X_m,Y_m)$ of the target  in frame $m$ relative to its position $X_0$ in the reference frame.
and used them to recover $S$ astrometric parameters $\xi_s$ of the target by fitting their change in time $t$ with the model
$\vec{\mathbf{\nu \xi + \Psi(t) = \Delta X}}$ \citep{Lazorenko2011},
which, in the general case, includes the orbital motion $\Psi(t)$ of the target
and  is solved as described by \citet{JS2013}.} For the field stars with  $\Psi(t)=0$, the  solution of 
 { this model provides the residuals $x_m$ and parameters  $\xi_s$ which refer to the system of  reference stars set by  Eq.~(\ref{eq:z}).
The correction terms to obtain their absolute values have to be determined using external data.}

\subsubsection{Deriving $R_{\rm opt}$ }{\label{Ropt}}

Eq.~(\ref{eq:balance}) is  solved numerically for $R_{\rm opt}$, provided that the coefficients  { $B$ and $b$ of the model expression } for $\sigma_{\rm atm}$ are known. These values, given in \citet{Lazorenko2006}, are approximate because of the differing observing conditions, and  the measured image motion $\sigma_{\rm atm}$ can contain additional components of instrumental origin \citepalias{Lazorenko2009}. Therefore we assumed that  the actual image motion  is  $z\sigma_{\rm atm}$, where $z$ is an amplitude factor. To find the value of $z$, we varied  it in the range of  $0.5 \ldots 2.5$, and derived $R_{\rm opt}$ by numerical solution of  Eq.~(\ref{eq:balance}).  Then we performed the astrometric reduction for   the target star (Sect.-s \ref{field_d},~ \ref{astr_mod_T}) and computed the estimator $\Upsilon(z)$, whose minimum indicates the best  value of $z$ (typically 0.8--1.5). {As suitable estimator, we used the index $ \Upsilon(z) = \sum_m (x_m/\sigma_m - x_{m-1}/\sigma_{m-1})^2$ of high-frequency noise in the positional  residuals of individual frames. For the final steps of the reduction, we used a similar expression but based on the high-frequency noise in the epoch residuals, where any orbital motion (if detected) was removed.}

\section{Results and mitigation of systematic errors}{\label{res}}
\subsection{Single-frame precision}{\label{fr_pr}}
The precision of a single frame measurement $\sigma_m$ depends on the photocentre measurement uncertainty, the atmospheric image motion, and uncertainties in the transformation to the reference frame. Our model of $\sigma_m$ \citepalias{Lazorenko2009} describes  the measured dispersion of residuals $x_m$ well, which we verified again by processing field stars as if they were the target, followed by the comparison of $\sigma_m$ and the measured r.m.s. of $x_m$.
\begin{figure}[htb]
\begin{tabular}{@{}c@{}}
\resizebox{\hsize}{!}{\includegraphics*[bb = 56 70 259 158, width=\linewidth]{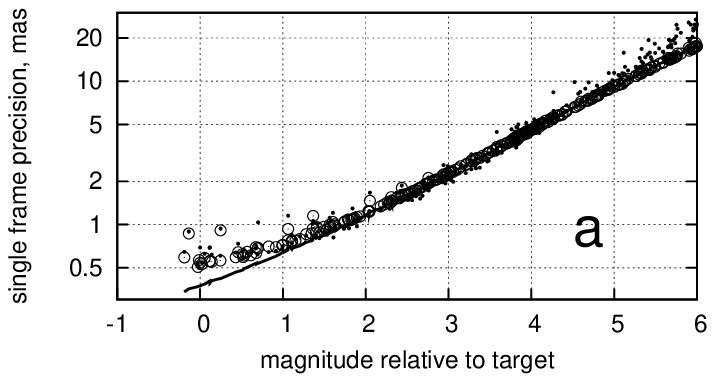}}\\
\resizebox{\hsize}{!}{\includegraphics*[bb = 56 50 259 158, width=\linewidth]{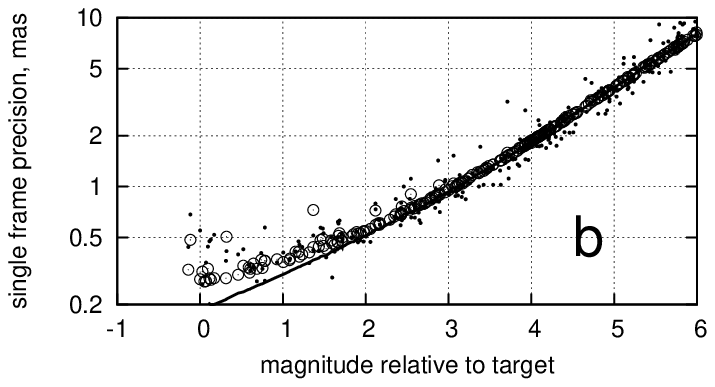}}\\
\end{tabular}
\caption{Single-frame precision as a function of magnitude for stars in the FoV with \dwfive.  Measured (dots) and model $\sigma_m$ (open circles) values are shown, and the solid line indicates the dominant photocentre uncertainty component for {\bf a}) the average of a full set of frames; {\bf b}) an epoch with excellent 0.38\arcsec\ seeing.  Magnitudes are given relative to \dwfive\ ($m_I=17.5$). }
\label{x_m}
\end{figure}

We  illustrate this and the next reduction steps with the help of the target Nr. 5 in Table \ref{ident} \citepalias{JS2013}, whose Simbad identifier is \object{DENIS-P J0716478-063037} and hereafter is referred to as \dwfive. For reference stars close to \dwfive, the comparison is shown in Fig.\,\ref{x_m}a, where only observations obtained in 0.4--0.8\arcsec\ seeing are plotted. This figure looks similar for all targets. The component that corresponds to the uncertainty of the photocentre measurement (solid line) is dominant except at the bright end, where the contribution from atmospheric image motion and reference frame noise become significant. Evidently, the model values $\sigma_m$ match  the observed r.m.s. of $x_m$ well over a range of the six magnitudes.

At the bright end, the single-frame precision is about 0.6 mas, which is higher than $\sigma_m=0.3$ mas obtained in the test study of the {\small FORS2 } astrometric performance \citep{Lazorenko2006}, but this is a consequence of a different observation strategy. In the current programme, we specifically restricted the exposure duration to avoid systematic errors caused by saturation in very good seeing (Sect.~\ref{obs}), whereas \citet{Lazorenko2006} aimed at  determining  the best astrometric precision with a single series of frames in stable seeing and with non-saturated target images with $1.5-2 \times 10^6$ electrons. The threefold gain in flux and a slightly better seeing (0.55\arcsec\ in \citealt{Lazorenko2006} compared with 0.6\arcsec here) are responsible for the difference in precision, which becomes negligible when scaled to similar conditions. Consequently, another investigation of the {\small  FORS2} astrometric precision during the timespan of five month \citepalias{Lazorenko2009} yielded the same $\sigma_m=0.6$\,mas precision for stars with $0.5-0.7 \times 10^6$ electrons in the image. For extremely bright stars in that study, the precision improved to 0.3--0.4\,mas, but such stars are not measured here because of the saturation problem.

For a single series of images obtained in very good seeing of 0.37--0.42\arcsec~(shown in Fig.\,\ref{x_m}b), the model prediction also  matches well except for the brightest stars, again, because of saturation. The impact of good seeing on the precision is evident with a twofold decrease of random { uncertainty} to 0.35~mas for bright stars. We conclude that the model value of $\sigma_m$ adequately represents the random { uncertainty}. This is an important prerequisite for separating noise and systematic components in the epoch residuals, because the random part decreases inversely to the square root of the number of exposures, and its value is well predicted by our model of $\sigma_m$.

\subsection{Epoch residuals}{\label{ep_res}}
By taking the averages of $n_e$ individual frame residuals $x_m$ available for each epoch $e$, we derived the epoch-averaged residuals $x_e$. These quantities are not directly used in the standard astrometric reduction (Sect.\ \ref{layout}), for example, to compute the astrometric parameters {\vec \xi}. However, they are critical to  search for and eliminate systematic errors with an inter-epoch variability. The epoch residuals take into account the individual measurement precisions
\begin{equation}
\label{eq:e0}
 x_e=\sum_{m \in e} x_m \sigma_m^{-2}/\sum_{m \in e} \sigma_m^{-2}, \qquad e=1,2 \ldots N_e,
\end{equation}
where $N_e$ is  number of epochs. They correspond to the usual weighted average of $n_e$ random normal values $X_m$,$Y_m$ that have precisions of
\begin{equation}
\label{eq:e01}
{\varepsilon}_e = \left(\sum_{m \in e} \sigma_{m}^{-2}\right )^{-1/2},
\end{equation}
which we call the nominal precision of the epoch residuals. This formal precision of the measured epoch data does not depend on the reduction that follows, in contrast to ${\sigma}_e$, which specifies the r.m.s. of the model-dependent fit residuals $x_e$ 
\begin{equation}
\label{eq:ap2}
\sigma_e^2=
\left. \sum_{m,m' \in e}^{n_e} \{ \vec{D- \boldsymbol{\nu} N^{-1}\boldsymbol{\nu^T})} \}_{m,m'}
	\sigma_m^{-2}\sigma_{m'}^{-2} \right /
      \left ( \sum_{m \in e}\sigma_m^{-2} \right ) ^{2},
\end{equation}
{ where {\boldmath $\vec{N }$} is the normal matrix \citepalias{Lazorenko2009}.}
The least-squares fit absorbs some noise from the measurements and induces correlations between the residuals $x_m$, therefore we have $\sigma_e < \varepsilon_e$. The equality is reached in the limit $N_e  \to \infty$, where Eq. (\ref{eq:ap2}) equals Eq. (\ref{eq:e01}).
{ Using  expression $E[(x_e/ {\sigma}_e)^2]=1$, where  the operator $E[]$ denotes the mathematical expectation,   } we can introduce the reduced  $\chi^2$ value
\begin{equation}
\label{eq:e1}
 \chi^2= (2N_e)^{-1}  \sum_e^{2N_e} (x_e/ {\sigma}_e)^2,
\end{equation}
whose deviation from unity indicates systematic errors or the signature of a companion. For the following statistical analysis, we have to know the degree of freedom (DoF) of $\chi^2$. By numerical simulations, where we used the actual observing conditions and added random noise to the measured data, we found that $ \sum_{e=1}^{2N_e} (x_e/ { \varepsilon_e})^2 = 2N_e-S+\Delta S$ (the sum involves $N_e$ data points both in RA and Dec) with $\Delta S= 0.39 \ldots 0.44$. This sum evaluates the effective number of DoF when all in-frame information is compressed to 2$\times$$N_e$ epoch residuals.  For our observations, $N_e=10.8$ on average, therefore $\mathrm{DoF} \simeq15$ if $S=7$. Another convenient representation is $\mathrm{DoF}=2\, \theta^2\, N_e$, where the value $\theta^2 \approx 0.70$ is typical for our programme.

When the data are free of systematic errors, the expectation of the r.m.s. of $x_e$ is $\sigma_e = \theta \, {\varepsilon}_e$, thus
\begin{equation}
\begin{array}{lcr}
\label{eq:euu}
  \langle (x_e/ {\sigma}_e)^2 \rangle =1; &
  \langle (x_e/ {\varepsilon}_e)^2 \rangle = \theta^2; &
  \sigma_e^2 = \theta^2 \varepsilon_e^2,  \\
\end{array}
\end{equation}
where the angle brackets indicate averaging over epochs.

\begin{figure}[htb]
\begin{tabular}{@{}c@{}}
\resizebox{\hsize}{!}{\includegraphics*[]{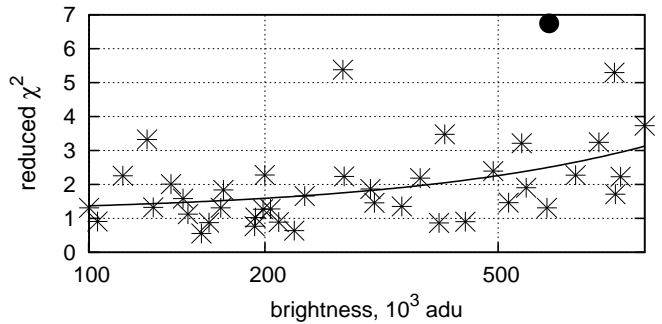}} \\
\end{tabular}
\caption {Reduced $\chi^2$ for bright stars in the \dwfive\ field (asterisks) as a function of brightness. The solid line corresponds to the best linear fit, and the target itself is marked with a filled circle. Data were reduced with $k=10$ and the chip's instability (Sect. \ref{chips}) was corrected.}
\label{xi_i}
\end{figure}

Systematic errors become apparent primarily by a substantial increase of $\chi^2$ for bright stars. We illustrate this in Fig. \ref{xi_i} with stars in the \dwfive\ field, which was observed in a wide range of seeing conditions. The preliminary reduction of reference stars processed as targets produced large epoch residuals characterised by $\chi^2 \gg 1$, showing a general tendency of increasing $\chi^2$ with the star brightness.

\begin{figure}[!ht]
\begin{tabular}{@{}c@{}}
\resizebox{\hsize}{!}{\includegraphics*[bb = 54 69 259 140]{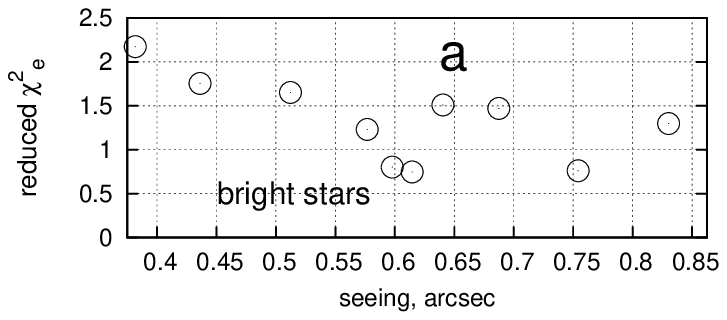}} \\
\resizebox{\hsize}{!}{\includegraphics*[bb = 54 69 259 140]{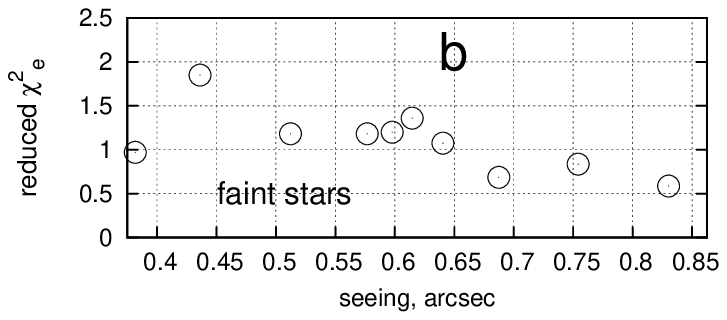}} \\
\resizebox{\hsize}{!}{\includegraphics*[bb = 54 69 259 140]{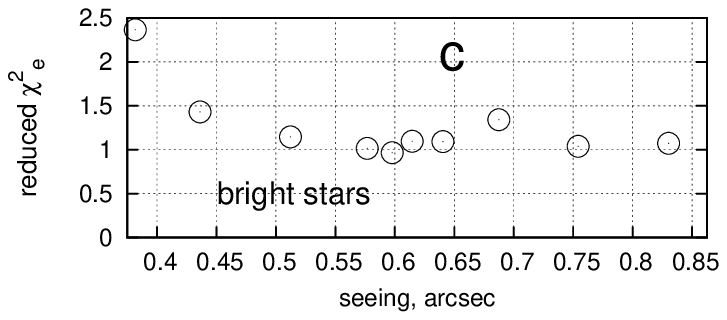}} \\
\resizebox{\hsize}{!}{\includegraphics*[bb = 54 50 259 140]{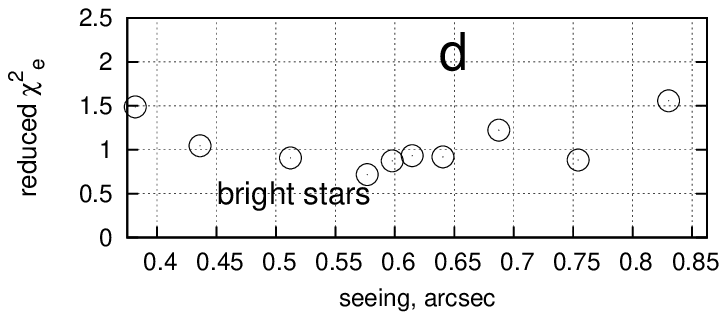}} \\
\end{tabular}
\caption {Dependence of the reduced $\chi^2_e$  on the epoch average seeing {\bf a)} for brightest field stars near \dwfive \, in the initial data with  corrected chip instability (Sect.\ref{chips});  {\bf b)} same for stars about 1~mag fainter; {\bf c)} after correction for space-correlated errors (Sect.\ref{cor}); {\bf d)} with all corrections of Sect.\ref{cor} and Sect.\ref{se_nc}. Standard computations with a reduction parameter $k=10$.
}
\label{xi_e}
\end{figure}

Systematic errors also give rise to a seeing-dependence of the quantity $\chi^2_e$,  which is the mean of $(x_e/ {\sigma}_e)^2$ for bright stars at the epoch $e$. As shown in Fig.\,\ref{xi_e}a, the $\chi^2_e$-values tend to be higher  in good seeing. This dependence is less pronounced for stars about one magnitude fainter (Fig.\,\ref{xi_e}b). This indicates that the amplitude of systematic errors is similar to $\sigma_e$ for bright stars, that is 0.05--0.2 mas. Therefore, the excess in $\chi^2_e$ is registered primarily for these stars and only in good seeing.
 
\subsection{Systematic errors in the epoch residuals}{\label{se}}
\subsubsection{Calibration files}{\label{cal}}
To investigate the origin of the systematic errors, we used the information provided by the positions of reference stars and performed a special calibration computation for each star near the field centre. These stars were processed as targets with $k=10$ and the actually targeted UCD was excluded from the reference objects. This provided us with a calibration file that has two important applications.

First, because { the majority of reference stars in our sample are very distant (there are only 20 reference stars that have a measured trigonometric parallax larger than 10 mas) }and any companion signature would be negligibly small, we can use the epoch residuals $x_{e, \rm rf}$ and $y_{e, \rm rf}$ of these stars for the detection, modelling, and mitigation of systematic errors.  Second, it becomes possible to evaluate and consequently minimise the statistical difference between the scatter of epoch residuals for reference stars and the target, which is necessary to find traces of a weak orbital signal in the latter.

\subsubsection{Instability of the relative CCD-chip position}
\begin{figure}[htb]
   \centering
\resizebox{\hsize}{!}{\includegraphics*[bb = 56 50 247 281]{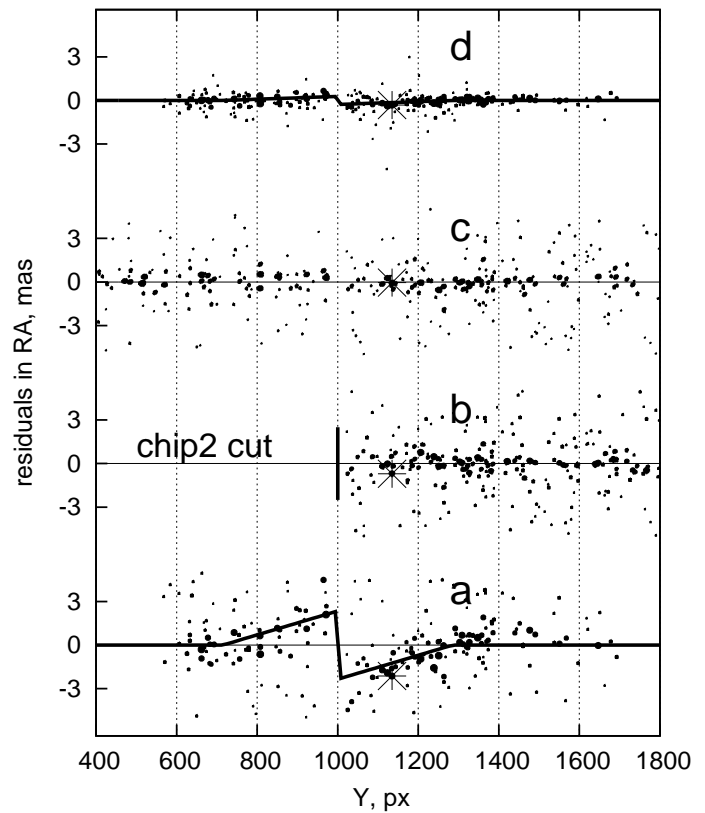}} \\
\caption {Positional residuals in RA on 24.11.2011 for stars in the \dwfive\ field computed with $k=10$ and $R_{\rm opt}=815$ px. {\bf a)} The original two-chip data showing the discontinuity at the chip edge with a full amplitude of ${g_x}=0.0376$~px, which is approximated by the function ${G_x(y)}$ (solid line); {\bf b)} only data of chip 1; {\bf c)} {after  offset of chip 2 photocentre positions in RA by ${g_x^*}$}; {\bf d)} an example of an epoch with good seeing and small discontinuity (${g_x}=0.0044$~px). The data corresponding to 
UCD are marked by an asterisk.}
\label{dw3chips}
\end{figure}

{\label{chips}}
The inspection of the epoch residuals $x_{e, \rm rf}$ and $y_{e, \rm rf}$ of stars in the calibration files sometimes revealed unexpected distribution patterns close to the border between the two chips making up the CCD detector of {\small FORS2}. A typical case is presented in Fig.\,\ref{dw3chips}a. A discontinuity between the residuals in chip 1 ($y>1000$~px) and chip 2 ($y<1000$~px) is present at the location of the inter-chip gap ($y=1000$). We described this systematic trend with a smooth function $G(y)$ that is anti-symmetric about the gap position. The discontinuity amplitude was usually registered at the noise level, but in many cases it was detected with a significance of 5--10 sigma, and could reach a peak-to-peak amplitude of $g=0.007-0.010$ px, either in RA or Dec or in both dimensions. The occurrences of large amplitudes appears randomly distributed over the programme calendar, except for the time interval of November 19--24,  2011, when unusually large discontinuities of equal sign in RA ($g=0.02-0.05$ px) were registered in five out of six epochs taken during this period. Surprisingly, the discontinuity was not seen in the epoch residuals of DENIS-P J0615493-010041 (hereafter \dwone) observed on 19 November 2011 only one hour before another target for which a large discontinuity amplitude was detected. This riddle will be resolved later in this section.

One of the extreme examples detected on 24 November 2011 in RA for \dwfive\ is shown in Fig. \ref{dw3chips}a and has a peak-to-peak amplitude of $g=0.0376 \pm 0.0030$~px~$=4.74 \pm 0.38$~mas. We found that the systematic behaviour of the epoch residuals as a function of $y$ disappears when discarding the data of one chip in the processing, see Fig.\ref{dw3chips}b which shows chip 1 residuals obtained without using data from chip 2.

We investigated the origin of this behaviour and found that it is not an artefact of the PSF modelling at the chip edges. Instead, it must be caused by the relative shift of either of the CCD chips (as solid-body motion) along the $x$ or $y$ axis with the corresponding effect in RA or Dec. The effect can be understood as follows: consider a target  star imaged on chip 1 just above the gap between the chips and let all images in chip 2 be shifted by $+g$ along the $x$-axis. Because half of the reference stars are displaced, this results in the change of the RA zero-point by $+g/2$, and we see a $-g/2$ change in the positional residuals of the target. In the same way, the positional residuals of a target imaged on chip 2 close to the gap are biased by $+g/2$. The effect amplitude decreases with increasing distance from the gap, because of the decreasing relative weight of the reference stars from the other chip. A piecewise linear function describes this systematic effect sufficiently well:
\begin{equation}                               
\begin{array}{rrl}                               
\label{eq:g}
   G_x(y)  =& 0.5\  g_x\ (1- \tilde{y})\hspace{2mm}\mathrm{if} &  y<1000   \\
   G_x(y)  = &-0.5\ g_x\ (1- \tilde{y})\hspace{2mm}\mathrm{if} &  y>1000   \\
   G_x(y)  = &0\hspace{2mm}\mathrm{if} &   |y-1000|>0.35\,R_{\rm opt},  
\end{array}
\end{equation}
where $\tilde{y}= |y-1000|/(0.35\,R_{\rm opt})$ is a normalised distance. The function (\ref{eq:g}) is valid in RA and involves one free parameter $g_x$, and a corresponding function depending on $g_y$ describes the {effect of chip motion along the Dec axis, that is orthogonal to the chip gap}. By fitting the calibration file data with the model (\ref{eq:g}), we derived the parameters $g_x$ and $g_y$ for every epoch and moved all chip 2 photocentre positions by constant offsets 
\begin{equation}                               
\begin{array}{ll}                               
\label{eq:g_}
   g_x^*  = g_x/\theta^2  \hspace{2mm}\mathrm{and} &       g_y^*  = g_y/\theta^2,
\end{array}
\end{equation}
which are slightly larger than $g_x$,  $g_y$. Here we considered that the values of $g$ are determined relative to the average chip position (i.e. including the epoch with a large shift), thus are biased. Using the term $1/\theta^2$, we approximately removed this bias in the zero-points, obtaining accurate shifts $g^*$. The dataset
\begin{equation}                               
\begin{array}{ll}                               
\label{eq:gg}
   X_{\rm corr.}  = X_{\rm meas.} -  g_x^*; \quad &    Y_{\rm corr.}  = Y_{\rm meas.} -  g_y^*,
\end{array}
\end{equation}
where the chip 2 motion was corrected (e.g.\ Fig.\,\ref{dw3chips}c) is free of the systematic pattern.

A potential origin of the chip motion was proposed by the ESO User Support Department (van den Ancker, 2013, priv. comm.). The {\small FORS2} camera has two interchangeable CCDs (red- and blue-optimised) of which only one is mounted at a given time. To exchange the CCD, the cryostat is warmed up and subsequently cooled down and the 
associated thermal shock can cause the instability in relative chip position at sub-micrometer level. This hypothesis is supported by that fact that we recorded epochs with large amplitudes of $g_x$ around times when the red CCD, which was used for our observations, was temporarily replaced, requiring cryostat interventions on November 4 and 25, 2011.

\begin{figure}[htb]
\resizebox{\hsize}{!}{\includegraphics*[bb = 55 326 221 479]{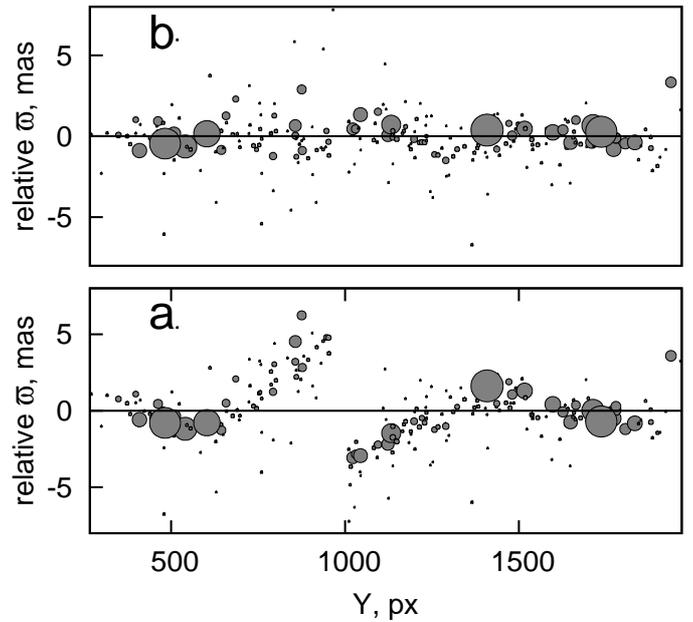}} \\
\caption {Systematic change of star parallaxes in the \dwone\ field. {\bf a)} Measured values with the bias pattern caused by the chip displacement; {\bf b)} The corrected data. The chip border is at $y=1000$~px and the dot size is proportional to the star brightness.}
\label{pi_dw1}
\end{figure}

As we noted above, the observations of \dwone\ on 19 November 2011 do not show the signature of a large relative chip shift, although this might be expected. Instead, we discovered a systematic change of the relative parallax values of field stars along the $y$-axis, that is similar in shape to the function $G$ with a peak-to-peak amplitude of 8~mas, see Fig.~\ref{pi_dw1}a. The chip instability is evidently the reason of this effect. Due to differential nature of the measurements, a chip displacement that changes in time proportional to the parallax factor cannot be distinguished from the parallax motion, and the measured parallaxes of field stars are biased with an amplitude that depends on $y$ (Fig.~\ref{pi_dw1}a). Consequently, the discontinuity in the epoch residuals disappears. To correct for the bias, we fitted the distribution of field star parallaxes with a function of the type (\ref{eq:g}) and then removed the systematic component, which yielded a bias-free distribution of parallaxes (Fig.~\ref{pi_dw1}b). Finally, a similar correlation can cause biases in the proper motions if $g$ changes linearly in time, and a corresponding correction procedure was applied to the proper motions, too.

\begin{figure}[htb]
   \centering
\resizebox{\hsize}{!}{\includegraphics*[bb = 55 50 224 140]{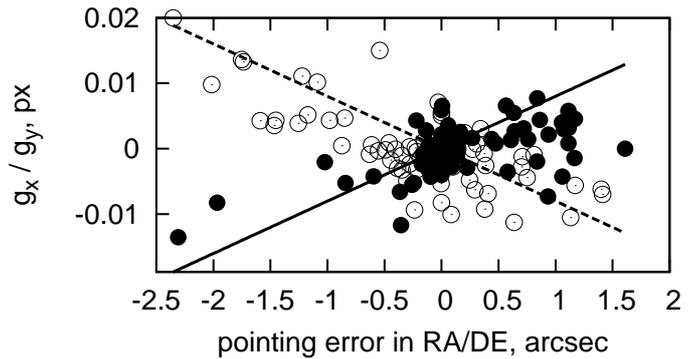}} \\
\caption {Correlation between the measured motion of chips ${ g}_x$ and the telescope pointing error in Dec (filled circles); between
${ g}_y$  and the pointing error in RA (open circles). Lines show the expected image shift due to the rotated mounting of chips.}
\label{rot}
\end{figure}

In addition to actual chip instability, the discontinuity in measured positions between chips can be caused be the slightly inclined mounting of chip 2 relative to chip 1 by the angle ${\bar \gamma}=0.083$\degr {(see ESO documentation\footnote{\url{ http://www.eso.org/sci/facilities/paranal/instruments/fors/doc}}). } If the telescope pointing is not perfect and deviates by $\Delta_x$, $\Delta_y$ from its 'normal' pointing, the star field in chip 1 is also moved by $\Delta_x$ and $\Delta_y$, but in chip 2 the images are moved by $\Delta_x+{\bar  \gamma} \Delta_y$ and $\Delta_y - {\bar \gamma} \Delta_x$.  This is equivalent to a chip 2 pseudo-displacement by ${\bar \gamma} \Delta_y$ and $-{\bar  \gamma} \Delta_x$.  For the VLT Unit Telescope 1, the pointing error $\Delta$ exceeded 1\arcsec\ in observations of seven targets for at least one epoch. In Fig. \ref{rot}, we compare the induced image shifts ${\bar \gamma} \Delta_y$ (dashed line) and  $-{\bar  \gamma} \Delta_x$ (straight line) with ${ g}_x$ and ${ g}_y$ derived from the calibration files for all available epochs of the seven targets. Considering that the precision of $g$ is 0.0005--0.002 milli-pixel (depending on the star density and seeing), we found a considerable but incomplete correlation between the induced and measured  shift of chips. Thus, the measured relative motion of  chips can be in part only caused by the telescope pointing error. In the problematic period of November 19--24, 2011, illustrated in Fig.\,\ref{dw3chips}, the telescope pointing was almost perfect, and the detected motion along the chip gap is real with an amplitude of $\sim$0.3--0.7 micrometer. Its absolute value, however, cannot be determined because of the relative nature of the astrometric reduction.

The chip instability can be very small but well-determined in good observing conditions (Figure \ref{dw3chips}d corresponds to ${g_x}=0.0044 \pm 0.0008$~px = $0.55 \pm 0.10$~mas). Therefore, we corrected all chip 2 positions irrespective of the effect amplitude $g$ by moving $X$ and ${Y}$ in chip 2 by ${ g^*}_x$ and ${g^*}_y$, respectively. The uncertainty of individual corrections according to Eq.~(\ref{eq:g}) for most field stars is smaller than 0.01--0.03~mas and hardly affects the final astrometric precision. In poor seeing, as displayed in Fig.\,\ref{dw3chips}a, the correction uncertainty can degrade to 0.1--0.5~mas for objects near the chip gap.

\subsubsection{Correlation between epoch residuals}{\label{c_c}}

The epoch residuals $x_{e, \rm rf}$,$y_{e, \rm rf}$ in calibration files were corrected for the chip instability (previous section), thus the largest deviations were filtered out before the following analysis of systematic errors. We found no significant correlation between the epoch residuals and the star colour, magnitude, proper motion, parallax, or seeing in the calibration file data.

To investigate the error distribution over the CCD, we introduced the value 
\begin{equation}
\label{eq:rho}
\hat \rho=
\langle x_e x_{e,\rm rf}/( \sigma_e \sigma_{e,{ \rm rf}}) \rangle, 
\end{equation}
which expresses the correlation between the epoch residuals $x_e$ and $x_{e,\rm rf}$ of the target and of a field star at distance $r$. The brackets denote the average over epochs.

\begin{figure}[htb]
   \centering
\resizebox{\hsize}{!}{\includegraphics*[bb = 55 50 265 146]{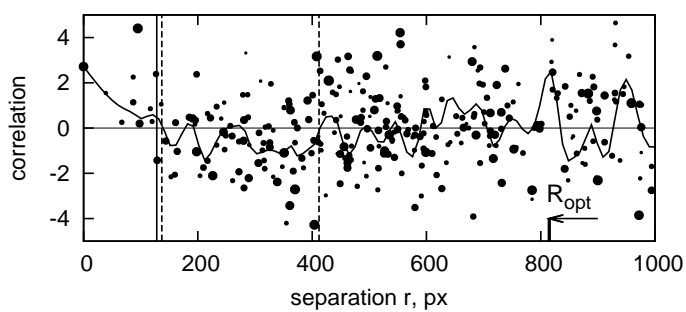}}
\caption {Individual $\hat \rho$ (dots) and smoothed  $\hat \rho(r)$ (solid curve) correlation between the epoch residuals of the target and reference stars in the calibration file of \dwfive\ as a function of separation $r$ with {positive (left of the vertical solid line) and negative (between dashed lines) loci.} The symbol size is proportional to the star brightness and \dwfive\ is located at $r=0$.}
\label{c1415}
\end{figure}

Fig.\,\ref{c1415} presents a typical distribution of $\hat \rho$. 
By { averaging} the $\hat \rho$ values, we obtained the function $\hat \rho(r)$, whose shape does not significantly depend on the choice of the star treated as the 'target'. {We found that $\hat \rho$  is usually {positive} on small scales from $r=0$ to $r\simeq 0.05-0.08 \,R_{\rm opt}$ and {negative} at larger distances } $r\simeq 0.1 \,R_{\rm opt} - 0.5 \,R_{\rm opt}$. 
 
A similar correlation is often seen in experimental data when the random signal limited in space (or time) is fitted by a polynomial. By numerical simulation, \citet{Andronov} and  \citet{Lazorenko1997} have shown that the remainders of the fitting (in our case, the epoch residuals obtained after polynomial fitting in CCD space) usually exhibit an oscillating  autocorrelation function, whose shape depends on the polynomial order and, to a minor extent, on the power spectrum of the {signal. Generally, the autocorrelation function of detrended measurements is positive at small scales before turning to zero at 0.1--0.15 times the 'normalised series length'. In one-dimensional space, this quantity is defined as a twice the series length ($2\,R_{\rm opt}$ for the present instance of reduction) divided by the polynomial order divided by two ($k^*/2$ in our case). Because $k^*=k/2-1$, the normalised series length is simply $R_{\rm opt}$ for the calibration file that was generated with $k=10$. The general properties of time series therefore predict the existence of the positive correlation at $r<$(0.1--0.15)$R_{\rm opt}$, and the negative correlation at larger $r$ up to $0.6 R_{\rm opt}$. The autocorrelation minimum occurs at $(0.25-0.35)R_{\rm opt}$. The two-dimensional astrometric data (Fig.\,\ref{c1415}) match the theoretical predictions for the autocorrelation shape, scales, and signs of de-trended measurements  well. Therefore, this correlation pattern is not specific to {\small FORS2} astrometry.
          
Correlations can also originate from high-order optical distortions and PSF changes across the CCD. Because they are related to the actual telescope alignment, these factors cause complicated image displacements that are not modelled by the basic functions $f(x,y)$, {but they tend to be stable during one epoch}. The difference  $\alpha_e$ between the real and the modelled photocentre positions is the uncompensated systematic position error, that produces the spatial correlation $\hat \rho$ in the epoch residuals.}

\subsubsection{Detection and inhibition of systematic errors}{\label{cor}}

The elimination of $\alpha_e$ for individual frames is problematic because of the {low signal-to-noise ratio (S/N) }of  $\alpha_e/\sigma_m$, but it becomes possible for an epoch, because of the higher S/N of $\alpha_e/\sigma_e$ in the epoch residuals. {For the computations, we used the sample estimates $\hat \rho$, but the direct estimates of  $\alpha_e$ are inapplicable, because they are too noisy (Fig.\,\ref{c1415}).  Instead of  $\alpha_e$ }, we calculated the quantities $\Delta_e$, which were defined to ensure minimum variance and low correlation in the corrected residuals $x_e$. We applied two different types of space-dependent corrections: $\Delta'_e$ for small-scale and $\Delta''_e$ for large-scale systematic errors.

For each epoch $e$, we averaged the values of $x_{e,\rm rf}$ for the reference stars in the calibration file encircled within the small radius $r_{\rm small}=\beta' R_{\rm opt}$ from the target star, which produced the averaged residuals $x_{e,\ast}$ {and their} precision $\sigma_{e,\ast}$. Using these values instead of $x_{e,\rm rf}$ and $\sigma_{e,\rm rf}$ in Eq. (\ref{eq:rho}), we computed $\hat \rho$, which now represents the correlation between the target and the nearby star group residuals. We searched for the strongest positive correlation $\hat \rho$ by varying $\beta'$, usually reached at $\beta'=0.05-0.15$. For \dwfive, {the maximum is at }$r_{\rm small}= 128$~px  $\simeq16\arcsec$, which is marked by a solid vertical line in Fig.\,\ref{c1415}. Using these data and the optimal subtraction of the common component in correlated values described in Appendix\,\ref{B2}, we computed the corrections $\Delta'_e$ and applied them to target positions $X_m$
at each frame of the epoch $e$.  Typical values of $\Delta'_e$ are 0.05--0.2~mas.

The precision $\phi_{\rm small}(e)$ of the correction $\Delta'_e$ depends on the {number} of nearby field stars and on the seeing. For low values $r_{\rm small}=0.1 \ldots 0.2\arcmin$, the number of encircled stars is low and for some field stars, $\Delta'_e$ cannot be found. It is important to recognise that despite apparently introducing additional noise due to the sometimes low nearby star number, the actual uncertainty of the corrected residuals does not increase, because of the optimal subtraction of the systematic error component. The method of removing the spatial correlation (Appendix\,\ref{B2}) is  a tool of decomposing the {observed} variance $D_e$ of $x_e$ into a sum (Eq.~\ref{eq:apB_1}) of the {model} variance $\sigma_e^2$  and the mean square of the systematic error $\alpha^2$. The systematic component is removed only partially to ensure the best final precision (Eq. \ref{eq:apB_4}). 

To search for systematic errors at large scales, we averaged the values of $x_e$ for reference stars in the calibration file within the annular area around the target between $r=0.6\, \beta\, R_{\rm opt}$  and $r=1.6\,\beta\,R_{\rm opt}$, where $\beta$ is {a variable}. The obtained values, which we denote as $\Delta''_e$, have opposite sign to $\Delta'_e$ because $\hat \rho$ is negative at large scales. Numerically, we found that the best convergence of the output epoch residuals is reached when $\Delta''_e$ is added to $X_m$ for $m \in e$.

While varying $\beta$, we processed each of $\sim$30 bright stars in the field centre as targets and applied the correction $\Delta''_e$ to the  photocentre positions $X_{m}$. The optimal value of $\beta$ is found when the sum of $x_e^2$, computed over the star sample, is minimum. For \dwfive, this value is $\beta =0.28$ and corresponds to the minimum of the function $\hat \rho$ at $0.28\,R_{\rm opt}$, which  agrees with its theoretical location at 0.25--0.35 times the 'normalised series length' \citep{Lazorenko1997}. Because $\beta$ is defined on large scales, it is applicable across the entire field, but $\beta'$ varies in space and therefore is specific to each target.

\begin{figure}[htb]
\resizebox{\hsize}{!}{\includegraphics*[bb = 56 50 260 159]{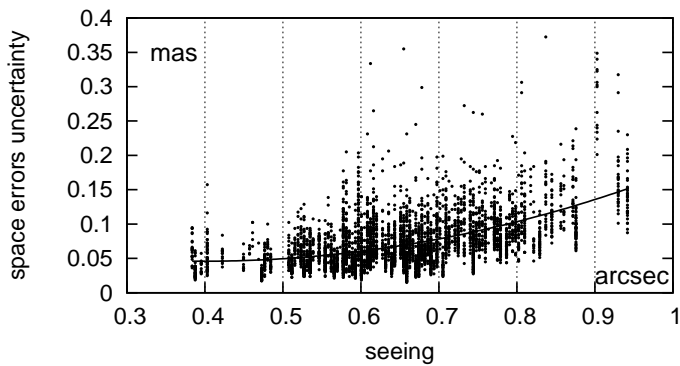}}
\caption {Dependence of the total uncertainty $\phi_{\rm space}(e) $ of spatial corrections on seeing for the brightest stars in all fields and for all epochs.}
\label{fiep}
\end{figure}

The astrometric reduction includes the complete computation of many bright field stars as targets {and the application of the corrections $\Delta'_e$  and $\Delta''_e$ with precisions $\phi_{\rm small}(e)$ and $\phi_{\rm large}(e)$, respectively, determined for each star individually. The sum $\phi^2_{\rm space}(e) =  \phi^2_{\rm small}(e) + \phi_{\rm large}(e)^2$ is the total variance of space-correlated errors in the reduced {\small FORS2} astrometry. The value of $\phi_{\rm space}(e)$ depends on seeing, as shown in Fig.~\ref{fiep}, because the systematic errors are better determined in good observing conditions  ($\sim$0.06~mas at average seeing). The variance  $\phi^2_{\rm space}(e)$ is added to the nominal variance $\varepsilon_e^2$ and appears in the covariance matrix $\vec{D}$ as the residual covariance of measurements in the frames $m$ and $m'$ of an epoch $e$. The related specificities in the reduction of correlated measurements are discussed in Appendix~\ref{B3}.}

\subsubsection{Residual systematic errors}{\label{se_nc}}
Despite the mitigation of systematic errors correlated in space, the $\chi^2$-value continues to depend on seeing (Fig.\,\ref{xi_e}c). Compared with the initial values (Fig.\,\ref{xi_e}a), the updated $\chi^2$-values are lower, but still unacceptably high in good seeing. One possible reason is the presence of an unknown systematic error ${\varphi}$ in the observations, which is not removed by the reduction and not revealed by the dependence of the residuals $x_e$ on observing conditions or model parameters. We modelled the corresponding effect in $\chi^2$ by the expression
\begin{equation}
\label{eq:fi}
   \chi^2  = 1+ 
	\langle \theta^2  \varphi^2 / \sigma_e^2\rangle,
\end{equation}
where the brackets denote the average over epochs, $\sigma_e^2$ is computed {taking into account} the uncertainties of space-dependent corrections $\phi^2_{\rm space}(e)$, and $\theta^2$ reflects the decrease of { uncertainty} by the least-squares fit.

By adjusting the $\chi^2$-values with the model Eq. (\ref{eq:fi}) for bright stars in the field centre, we derived $\varphi$, whose value is 0.08~mas for \dwfive\ and 0.03--0.14~mas for other targets with a median of 0.09~mas. The $\varphi$ estimate was derived assuming that its value does not depend on epoch and location in the FoV. Because of its typical magnitude of 0.1~mas (Table~\ref{t}), $\varphi$ dominates the space-dependent components and limits the precision of bright star observations, for which $\varepsilon_e$ is of similar amplitude. No clear dependence of $\varphi$ on observing conditions or field star density was found, except a weak upward trend {for crowded fields (Nr. 17 and 18), where photocentres are biased by nearby star images, and in fields with low star density.

The revised variance of systematic errors is now 
\begin{equation}
\label{eq:upd}
	\phi^2_e =   \phi^2_{\rm space}(e) +  \varphi^2,
\end{equation}
therefore} the nominal variance of epoch measurements is 
\begin{equation}
\label{eq:ver}
\psi_e^2=\varepsilon_e^2+\phi^2_e 
\end{equation}
{instead of $\varepsilon_e^2$.} Consequently, Eq. (\ref{eq:euu}) is updated to 
\begin{equation}
\begin{array}{lcr} 
\label{eq:corfin}
 \langle (x_e/ {\sigma}_e)^2 \rangle =1; & 
  \langle (x_e/ {\psi}_e)^2 \rangle = \theta^2; &
  \sigma_e^2 = \theta^2 \psi_e^2,  \\
\end{array}
\end{equation}
where $\sigma_e^2$ is defined by Eq. (\ref{eq:ap2}),  provided that computations are made while allowing for all systematic errors.

The quantity $\varphi$ is used to adjust the model expectation of epoch residuals to their measured r.m.s. The efficiency of including $\varphi$  in the reduction model is illustrated by Fig.\,\ref{xi_e}d, where $\chi^2_e$ is independent of seeing. In addition, it allows us to update the covariance matrix $\vec D$ (Appendix \ref{B3}).

\subsubsection{Seeing variability and systematic  errors}{\label{fwhm}}

Finally, we inspect the relation between seeing and intrinsic systematic space-dependent errors in the epoch residuals, which were not corrected with the approach of Sect. \ref{cor}. The definition of their magnitude $\alpha_{\rm syst}$ is somewhat ambiguous but it seems natural to specify it as the average squared sum of the small- and large-scale components $\alpha_e$  and $\Delta''_e$.

The characteristics of  $\alpha_{\rm syst}$ for {\small FORS2} were already analysed in \citetalias{Lazorenko2009}, following a slightly different definition, for a single sky field and five epoch observations with monthly spacing. There, the typical amplitude of $\alpha_{\rm syst}$ was found to be 0.16~mas within $1\arcmin$ from the field centre. For observations in a more restrictive 0.47--0.78\arcsec seeing range, which included 80\% of all images, they decreased to about 0.05~mas at the field centre and to 0.1~mas in the periphery.

\begin{figure}[htb]
\resizebox{\hsize}{!}{\includegraphics*[bb = 56 50 260 157]{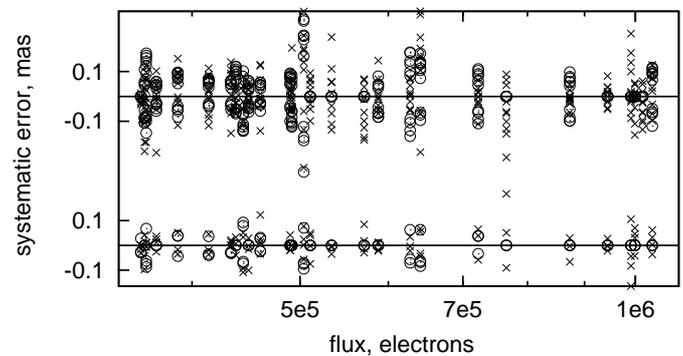}} \\
\caption {{Space-dependent errors in RA for each epoch: $\alpha_e $ (open circles) and   $\Delta''_e$ (crosses) as a function of star brightness in the \dwfourt\ field} for the complete ({top panel}) and  restricted ($0.47\arcsec < \mathrm{FWHM}< 0.78\arcsec$, {bottom panel}) datasets.
}
\label{dw14}
\end{figure}

Here, a similar estimate $\alpha_{\rm syst} \simeq 0.15$~mas  was obtained for a few sky fields with roughly the same 20\% ratio of non-standard seeing. By simulating restricted observing conditions and artificially rejecting frames obtained outside of 0.47--0.78\arcsec seeing, this value decreased to 0.09--0.11~mas. In both cases, the current estimates are close to those obtained in \citetalias{Lazorenko2009}. 

The effect of seeing restriction is illustrated by Fig.\,\ref{dw14}, which shows the components $\alpha_e $ and $\Delta''_e$  separately for bright stars in the \dwfourt\ field. The observations of this target stand out because of the high percentage (34\%) of non-standard seeing and the large systematic errors (Fig.\,\ref{dw14} top). When the astrometric reduction is made with the restriction $0.47\arcsec < \mathrm{FWHM}< 0.78\arcsec$, the error components become smaller (Fig.\,\ref{dw14} bottom, where only sufficiently well measured systematic errors are shown). When averaged over the complete target sample, however, the overall improvement by restricting the seeing is small ($\sim$10\%).

\begin{figure}[htb]
{\includegraphics*[bb = 56 50 263 156, width=\linewidth]{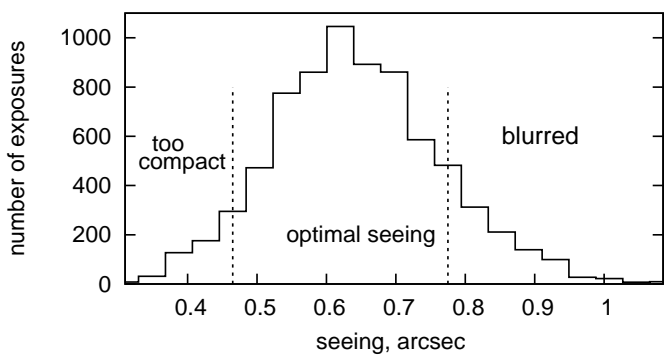}}
\caption{Distribution of actual seeing in {7442 frames collected for our programme between 26 November 2010 and 15 March 2013, and for which we measured the star photocentres.}}
\label{hist}
\end{figure}

A substantial 20\% fraction of the exposures taken for our programme was obtained outside the optimal seeing domain (Fig.\,\ref{hist}). In 45 epochs, most images were obtained with FWHM outside of the best 0.47--0.78\arcsec range, and in 12 epochs, all  images were obtained out of that restriction. For six targets, substantial ($>$50\%) occurrence of non-optimal seeing  was registered in three or more epochs. Discarding data using the seeing criterion is thus unacceptable, because is leads to a dramatic loss in number of observations that is not compensated for by a gain in precision.

\subsection{Adjusting $\chi^2$ of epoch residuals for field stars}{\label{conc}}
After removing the systematic errors, the $\chi^2_{\rm rf}$-values of epoch residuals (Eq. (\ref{eq:e1}))  for reference stars hardly depend on seeing anymore and are close to unity in most fields. Their dependence on brightness is also negligible, therefore our model of observational errors is now adequate. The following analyses revealed minor deviations of the $\chi^2_{\rm rf}$-distribution from the theoretical $\chi^2$-distribution,  however.

\subsubsection{Statistics of $\chi^2$}
\begin{figure}[htb]
\begin{tabular}{@{}c@{}}
\resizebox{\hsize}{!}{\includegraphics*[bb = 54 55 267 160]{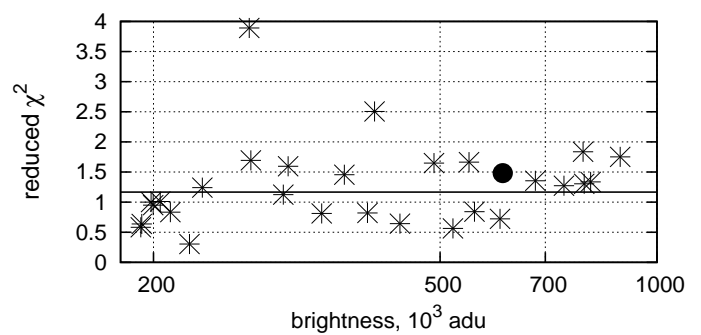}} \\
\end{tabular}
\caption {  {$\chi^2_{\rm rf}$  for bright reference stars (asterisks) and the target (black dot) in the  \dwfive\ field after removing systematic errors (reduction with $k=10$). }}
\label{xi_i_fin}
\end{figure}

{\label{av_conc}}
A typical $\chi^2_{\rm rf}$-distribution for field stars (each processed as a target object with $k=10$) is illustrated by Fig.\,\ref{xi_i_fin}. The individual $\chi^2_{\rm rf}$-values are concentrated around the average value $c_{\chi^2}^2$ close to unity  with a scatter, which ideally should match the theoretical $\chi^2$ distribution with DoF = 15 on average (Sect. \ref{ep_res}). The residuals $\chi^2_{\rm rf}-c_{\chi^2}^2 $ are expected to show an r.m.s. scatter of $\sqrt{2/DOF}=0.35$ and the probability of detecting deviations above three sigma, i.e.\ $\chi^2_{\rm rf}>c_{\chi^2}^2+1.1 $, should be low. 

The observed distribution of $\chi^2_{\rm rf}$, however, has a long non-Gaussian tail extending to extremely high values of 3--20 (one deviation in Fig.\,\ref{xi_i_fin}), which is incompatible with the statistical prediction. We verified that most outliers are related to non-standard cases such as elongated images, { apparent} binarity of the object, a close background star, CCD defects, and too large or too small FWHM, {and those were rejected by requiring $\chi^2_{\rm rf}<3$.}

\begin{table}[tbh]
{\small
\caption{Data related to the computation of $c^2_{\chi^2}$, {the residual { systematic} error $\varphi$, the r.m.s. of the epoch residuals $x_e$ and the epoch precision $\sigma_e$ for bright stars. The field number is given in column 1.} }
\centering
\begin{tabular}{@{}ccccccccc@{}}
\hline
\hline
Nr  & $\chi^2_{\rm rf}$  & $\chi^2_{\rm rf}$  & $c_{\chi^2}^2$ & $\chi^2_{\rm dw}$ & $\chi^2_{\rm dw}$ & $\varphi$&r.m.s.     &$\sigma_e$  \\
    &  faint             & br.               &                &  meas.            & corr.             &   mas     & mas       &  mas      \\
\hline
1  &   1.18              & 1.18              & 1.18           &      1.47          &  1.24            & 0.11      &  0.13     &      0.12   \rule{0pt}{11pt}\\
2  &   1.17              & 1.17              & 1.17           &      111           &   95\tablefootmark{(a)}             & 0.13      &  0.22     &      0.19   \\
3  &   1.12              & 1.12              & 1.12           &      0.52          &  0.46            & 0.12      &  0.14     &      0.13   \\
4  &   1.23              & 1.23              & 1.23           &      0.78          &  0.64            & 0.06      &  0.13     &      0.14   \\
5  &   1.18              & 1.18              & 1.18           &      0.89          &  0.76            & 0.08      &  0.11     &      0.11   \\
6  &   1.09              & 1.09              & 1.09           &      1.00          &  0.92            & 0.07      &  0.10     &      0.14   \\
7  &   1.11              & 1.11              & 1.11           &      0.87          &  0.78            & 0.09      &  0.12     &      0.13   \\
8  &   1.45              & 1.45              & 1.45           &      1.20          &  0.83            & 0.07      &  0.11     &      0.10   \\
9  &   1.49              & 1.49              & 1.49           &       122          &    82\tablefootmark{(b)}            & 0.11      &  0.16     &      0.14   \\
10 &   1.14              & 1.14              & 1.14           &      0.88          &  0.78            & 0.08      &  0.12     &      0.11   \\
11 &   0.99              & 1.37              & 1.03           &      1.01          &  0.98            & 0.10      &  0.12     &      0.12   \\
12 &   1.25              & 1.43              & 1.48           &      1.98          &  1.33            & 0.08      &  0.13     &      0.12   \\
13 &   1.21              & 1.21              & 1.21           &      2.22          &  1.84            & 0.07      &  0.14     &      0.15   \\
14 &   1.12              & 1.31              & 1.22           &      2.18          &  1.79            & 0.07      &  0.10     &      0.12   \\
15 &   1.09              & 1.09              & 1.09           &      0.93          &  0.86            & 0.06      &  0.12     &      0.11   \\
16 &   0.71              & 0.99              & 1.12           &      1.46          &  1.30            & 0.09      &  0.11     &      0.13   \\
17 &   1.44              & 1.57              & 1.54           &      1.37          &  0.89            & 0.14      &  0.16     &      0.15   \\
18 &   1.36              & 1.36              & 1.36           &      0.87          &  0.64            & 0.14      &  0.14     &      0.14   \\
19 &   1.10              & 1.28              & 1.29           &      2.42          &  1.87            & 0.10      &  0.16     &      0.17   \\
20 &   1.89              & 1.89              & 1.89           &      2.40          &  1.27            & 0.03      &  0.10     &      0.08   \\
\hline
\end{tabular}
\tablefoot{ \tablefoottext{a}{Target with  a strong signal from a companion that is not yet characterised.}
\tablefoottext{b} {Target with a 28 Jupiter mass  companion  \citep{JS2013}. }
}
\label{t}
}
\end{table}

The data show that $c_{\chi^2}^2$ can be described by first a constant and then, above a threshold brightness $I>I_0$, by a linear function: 
\begin{equation}
\label{eq:d}
c_{\chi^2}^2(I) = \left \{ \begin{array}{ll}
A, \hspace{2mm} &\mathrm{for}\hspace{2mm} I<I_0   \\
A +BI \hspace{2mm}&\mathrm{for}\hspace{2mm}  I>I_0, \\
\end{array} \right.
\end{equation}
with a small positive or,  in most cases, insignificant slope $B$. The data is listed in Table\,\ref{t} with $\chi^2_{\rm rf}$ given for the fainter (col.2) and brighter (col.3) half of $\sim$35 of the brightest stars near the field centre, $c_{\chi^2}^2$ at the target brightness computed with Eq. (\ref{eq:d}), and the measured $\chi^2_{\rm dw}$ for the targets. While these data refer to the reduction mode $k=10$,  $\chi^2_{\rm dw}$ for targets are given for the reduction that incorporates all modes $k$ from 4 to 16, because it better serves the purpose of the analysis. For the brightest stars, $\chi^2_{\rm rf}$  is  normally within 1.0 and 1.3, but in dense {fields (Nr. 17 and 20), it increases to 1.6--1.9. In the example of Fig.\,\ref{xi_i_fin},  the distribution of $\chi^2_{\rm rf}$ for reference stars is  fitted by a constant $c_{\chi^2}^2= 1.178$. For \dwfive, $\chi^2_{\rm dw}= 1.48$, which is well within  an r.m.s. of 0.49 of the data points} (in Table \ref{t}, $\chi^2_{\rm dw}= 0.892$ refers to the mean of $k=4 \ldots 16$ reduction modes; it also lies within the statistical scatter).

Distant reference stars have no or a negligible orbital signature and the {expected value }of $\chi^2_{\rm rf}$ is exactly 1.0. Therefore, the reference stars provide the calibration
\begin{equation}
\label{eq:ddc}
\hat {\chi}^2_{\rm rf} = \chi^2_{\rm rf} /  c_{\chi^2}^2(I), 
\end{equation}
which leads to a better estimate of $\chi^2_{\rm }$.

\begin{figure}[htb]
   \centering
\begin{tabular}{@{}c@{}}
\resizebox{\hsize}{!}{\includegraphics*[bb = 56 59 263 160]{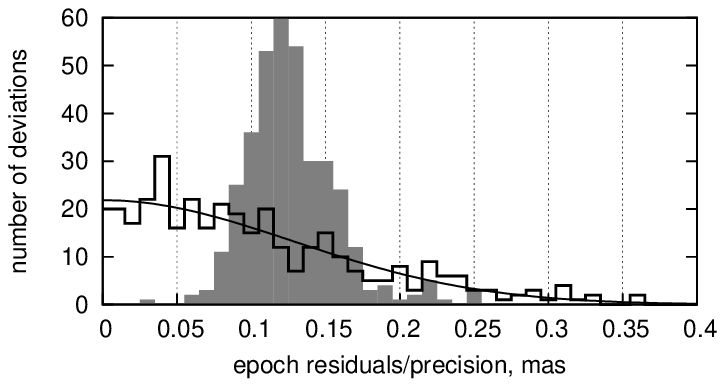}} \\
\resizebox{\hsize}{!}{\includegraphics*[bb = 56 49 263 160]{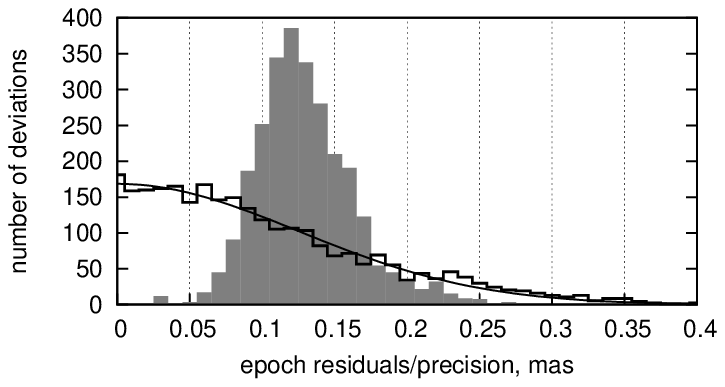}} \\
\end{tabular}
\caption {{\bf Histogram of the absolute values of epoch residuals (open histogram), the best-fit normal function  (solid curve), and the histogram of the epoch precision $\sigma_e$ (filled histogram) for epochs with 32 exposures and 0.6\arcsec\ seeing: for bright stars ({\it lower panel}) and UCDs only ({\it upper panel}).}
}
\label{histe}
\end{figure}

Considering that $\chi^2_{\rm rf}$ is the mean ratio of $x_e^2/\sigma^2_{e}$, one might suggest that the excess in $\chi^2_{\rm rf}$ detected for most fields (Table~\ref{t}) is due to systematically underestimated model values of $\sigma_{e}$. However,  for stars whose brightnesses are similar to the target within $\pm 0.5$~mag, the measured r.m.s. of  $x_e$ does not exceed $\sigma_{e}$ on average. This conclusion is based on the estimates given in the last two columns  of Table~\ref{t}, which correspond to observations in standard conditions with median seeing { (expressed by FWHM) of 0.6\arcsec\ and 32 frames.  These estimates were obtained by scaling the measured r.m.s. of  $x_e$ and $\sigma_{e}$ with the factor 0.6\arcsec/FWHM$*\sqrt{n_e/32}$, which makes the data more homogeneous. } Evidently, $x_e  \approx \sigma_{e}$ on average.

The histogram of the epoch residuals { (normalised to the standard seeing and frame number)} for bright stars within 50\arcsec\ from the target is shown in lower panel of Fig. \ref{histe} and is fitted by a normal distribution with sigma parameter $0.125 \pm 0.003$~mas, which is close to the independently computed  mean-square value of $0.126$~mas for $x_e$. For $\sigma_{e}$, whose histogram is also shown in the lower panel of Fig. \ref{histe}, we obtained the median of $0.126$~mas and the mean-square value of $0.128$~mas. We conclude that the model value  $\sigma^2_{e}$ matches the observed $x_e^2$ well, therefore the excess in $\chi^2_{\rm rf}$ has to have another origin. Note that while $\sigma^2_{e}$ refers to the epoch fit residuals, the nominal precision of epoch measurements $\psi_e$ is a factor $1/\theta=1.20$ higher, that is, 0.15~mas.

The interpretation of the excess in $\chi^2_{\rm rf}$ is difficult because of its low value. Considering the above analysis, we relate it to the uncertain determination of  $\sigma^2_{e}$. In other words, while we verified that the real and model $\sigma^2_{e}$ values are equal on average, the predictions for individual epochs can deviate from their actual values. This systematically biases the ratio $x_e^2/\sigma^2_{e}$ towards higher values, while $\sigma^2_{e}$ matches $x_e^2$ on average. Therefore, Eq.~(\ref{eq:ddc}) leads to better estimates of $\chi^2$, but a similar correction is not applicable to $\sigma^2_{e}$.

Finally, we estimated the precision of the epoch residuals for the data filtered by the seeing criterion, that is, using images with 0.47\arcsec < FWHM< 0.78\arcsec\ only, and obtained  0.11~mas for an r.m.s. of epoch residuals  and 0.13~mas for the nominal epoch precision, with only 10\% improvement compared with the complete data set.

\subsubsection{Cumulative distribution of $\chi^2$ for epoch residuals.}
\begin{figure}[htb]
\begin{tabular}{@{}c@{}}
\resizebox{\hsize}{!}{\includegraphics*[bb = 56 76 262 158, width=\linewidth]{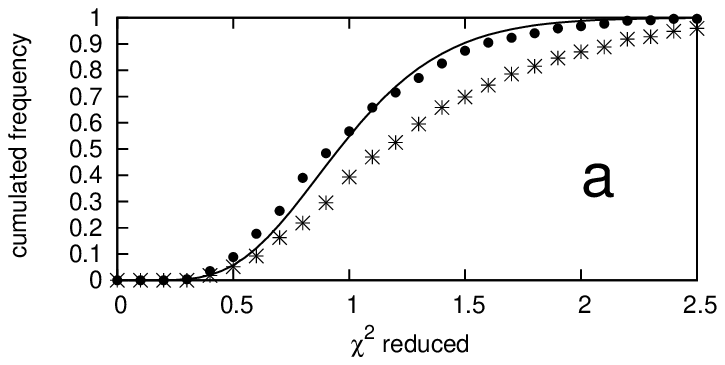}} \\
\resizebox{\hsize}{!}{\includegraphics*[bb = 56 55 262 158, width=\linewidth]{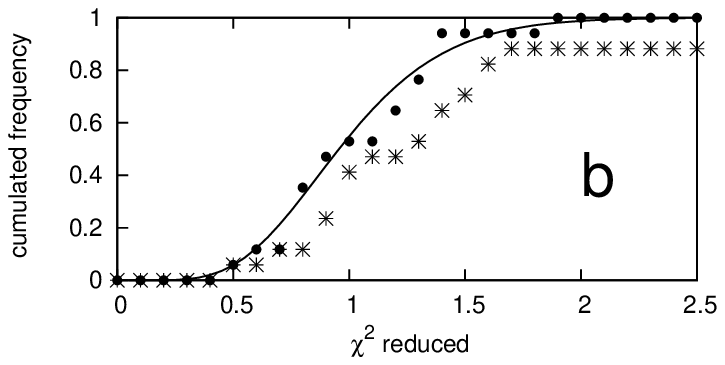}} \\
\end{tabular}
\caption{{ Normalised} cumulative distribution of the measured $\chi^2_{\rm rf}$ for epoch residuals (asterisks), theoretical distribution   $\chi^2(15)$ (solid line), and $\hat {\chi}^2_{\rm rf}$ after applying the correcting factor $c_{\chi^2}$ (filled circles)  {\bf a)} for bright field stars and {\bf b)} - for targets only.
}                                            
\label{h0}
\end{figure}

In Fig.\,\ref{h0}a, the measured cumulative distribution of $\chi^2_{\rm rf}$ for epoch residuals of field stars is compared with the theoretical $\chi^2$-distribution with DoF=15. The measured distribution significantly deviates from its theoretical expectation, but after the correction Eq.~(\ref{eq:ddc}) (filled circles), {they become compatible}. The same quantities but for targets only are displayed in Fig.\,\ref{h0}b. In summary, we showed that the statistics of the epoch residuals expressed in terms of $\chi^2$ follows the expected normal distribution sufficiently well.

\subsection{Astrometric precision for UCDs}

\begin{figure}[htb]
   \centering
\resizebox{\hsize}{!}{\includegraphics*[bb = 55 49 259 159]{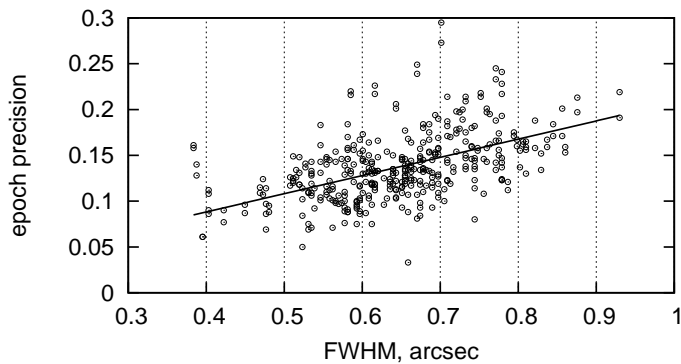}} 
\caption{Precision of the epoch residuals for UCDs for a frame series with 32 exposures.}
\label{sig_dw}
\end{figure}

{ The histogram of the epoch residuals for bright stars can be compared with that for UCDs only (upper panel of Fig.~\ref{histe}), which is fitted by a normal distribution with sigma parameter $0.128 \pm 0.009$~mas,  close to the  mean-square value of $0.137$~mas for $x_e$.  For $\sigma_{e}$, we obtained the median of $0.122$~mas and the mean-square value of $0.127$~mas. The derived values are very near to the corresponding estimates for bright stars; the precision characteristic of UCDs and bright stars is therefore identical. Note that these estimates correspond to FWHM=0.6\arcsec  and $n_e=32$ frames. If the precision parameters are not normalised to these standard conditions (e.g. frame series with small and large frame number are used with equal weights), the r.m.s. dispersion of the epoch residuals increases to  0.146~mas, which is the global astrometric precision of the programme \citepalias{JS2013}.

 Fig.~\ref{sig_dw} illustrates that the precision of the epoch residuals for UCDs (normalised to $n_e=32$ exposures) changes  from 0.1  to 0.2~mas almost linearly with FWHM.}

\begin{table*}[tbh]
\caption [] {Catalogue characteristics}
{\small
\centering
\begin{tabular}{@{}rcccccccccccc@{}}
\hline
\hline
Nr&DENIS-P &All           & Bright  &$R_{\rm opt}$&  $I_1$& $I_2$ &$\sigma_{\alpha,\delta}$ &$\sigma_{\mu}$&$\sigma_{\varpi}$& common   & Pix. scale              & (O$-$C)\tablefootmark{c}    \rule{0pt}{11pt}         \\
   &        &stars  &stars\tablefootmark{a} & (\arcsec)  &  (mag)  & (mag)   &   (\arcsec)                & (mas/yr) &  (mas)               &stars\tablefootmark{b}&   (mas/px)  & r.m.s. (\arcsec) \\
\hline
 1& J0615493-010041   &  245  &   23   &      120    &  16.1 &  22.4 & 0.062                   &        0.111 &  0.125       &    50    & $126.17\pm 0.13$   &  0.185             \rule{0pt}{11pt}\\
 2& J0630014-184014   &  178  &    8   &      112    &  15.1 &  22.1 & 0.053                   &        0.064 &  0.092       &    55    & $126.20\pm 0.11$   &  0.199          \\
 3& J0644143-284141   &  172  &   10   &      148    &  16.1 &  22.0 & 0.070                   &        0.118 &  0.110       &    41    & $126.10\pm 0.14$   &  0.246          \\
 4& J0652197-253450   &  135  &   15   &       88    &  14.5 &  20.4 & 0.059                   &        0.113 &  0.095       &    64    & $126.08\pm 0.12$   &  0.182          \\
 5& J0716478-063037   &  468  &   20   &      127    &  17.3 &  24.0 & 0.068                   &        0.114 &  0.112       &    50    & $126.45\pm 0.14$   &  0.190          \\
 6& J0751164-253043   &  431  &   23   &       77    &  16.2 &  21.6 & 0.073                   &        0.101 &  0.098       &    65    & $126.24\pm 0.12$   &  0.192          \\
 7& J0805110-315811   &  473  &   39   &       62    &  15.7 &  21.6 & 0.064                   &        0.101 &  0.110       &    65    & $126.07\pm 0.11$   &  0.202          \\
 8& J0812316-244442   &  457  &   23   &       91    &  16.3 &  22.9 & 0.045                   &        0.099 &  0.111       &    65    & $126.01\pm 0.09$   &  0.179          \\
 9& J0823031-491201   &  635  &   30   &      108    &  16.6 &  23.2 & 0.088                   &        0.064 &  0.106       &    62    & $126.26\pm 0.13$   &  0.239          \\
10&  J0828343-130919  &  156  &    7   &      142    &  15.4 &  22.1 & 0.042                   &        0.091 &  0.120       &    48    & $126.18\pm 0.09$   &  0.135          \\
11&  J1048278-525418  &  708  &   58   &       79    &  15.8 &  22.0 & 0.055                   &        0.119 &  0.094       &   125    & $126.01\pm 0.08$   &  0.178          \\
12&  J1157480-484442  &  406  &   27   &      122    &  16.0 &  22.4 & 0.075                   &        0.099 &  0.075       &    77    & $126.23\pm 0.11$   &  0.262          \\
13&  J1159274-524718  &  298  &    7   &       93    &  13.8 &  20.3 & 0.040                   &        0.116 &  0.107       &   116    & $126.32\pm 0.06$   &  0.181          \\
14&  J1253108-570924  &  599  &   54   &       43    &  15.8 &  22.0 & 0.073                   &        0.102 &  0.083       &    99    & $126.00\pm 0.10$   &  0.217          \\
15&  J1520022-442242  &  522  &   36   &       63    &  15.2 &  21.7 & 0.061                   &        0.106 &  0.132       &   109    & $126.08\pm 0.09$   &  0.206          \\
16&  J1705474-544151  & 1483  &   56   &       51    &  15.7 &  22.3 & 0.051                   &        0.107 &  0.108       &   144    & $126.12\pm 0.07$   &  0.179          \\
17&  J1733423-165449  & 1915  &   76   &       44    &  16.1 &  21.4 & 0.043                   &        0.125 &  0.112       &   118    & $126.46\pm 0.09$   &  0.180          \\
18&  J1745346-164053  & 1892  &   67   &       45    &  15.5 &  21.4 & 0.113                   &        0.121 &  0.108       &    50    & $126.09\pm 0.26$   &  0.272          \\
19&  J1756296-451822  &  790  &   25   &       45    &  14.3 &  20.5 & 0.074                   &        0.135 &  0.092       &    84    & $125.97\pm 0.12$   &  0.239          \\
20&  J1756561-480509  &  980  &   53   &       98    &  15.8 &  22.2 & 0.089                   &        0.069 &  0.054       &    82    & $126.30\pm 0.13$   &  0.251          \\
\hline
\end{tabular}
\tablefoot{{$\sigma_{\alpha,\delta}$ is the uncertainty of ICRF positions for stars identified with 'USNO-B';  $\sigma_{\mu}$ and $\sigma_{\varpi}$ are given for brightest stars.} \\
\tablefoottext{a}{ Number of stars within $\pm$0.5~mag from the target magnitude. }
\tablefoottext{b} {Number of stars in common with 'USNO-B'. }
\tablefoottext{c} {R.m.s. of the residuals 'FORS2'-'USNO-B'.}
}
\label{scale}
}
\end{table*}

\section{Catalogue of field stars}{\label{cat}}
The {\small FORS2} data collected for our project contain rich information on the kinematics and distances of a large number of faint field stars. We make this data available by publishing a catalogue containing 12\,000 stars with $I=16-21$~mag in 20 fields. The catalogue includes all stars within 1\arcmin 50\arcsec\ distance from the respective target (1\arcmin 30\arcsec\ for targets Nr.\,17 and 18, which have a high star density) whose photocentres were measured with acceptable precision. Elongated images (e.g. galaxies) and images of close star pairs were  rejected. Computations were restricted to the reduction mode $k=10$ only and reproduce those made for the calibration file (Sect. \ref{cal}), but for a much larger star sample and with all systematic corrections applied. This produced a complete set of astrometric parameters, including CCD coordinates at  the MJD=55700 epoch adopted for all reference frames. Positions were determined with precisions of $\sim$0.05--0.07~mas for bright stars. However, because they are given in the {local} system of reference frames with uncertain orientation and deformation caused by the telescope optics, they are not  directly available for external use. In our catalogue, these coordinates are transformed to ICRF and J2000.0 using external catalogues, unfortunately, with a loss of three orders of magnitude in precision.

\subsection{Transformation to the ICRF system}{\label{ICRF}}
The conversion from relative {\small FORS2} positions to the ICRF system was made by aligning field star coordinates with positions in the USNO-B catalogue \citep{USNO}. In each field, we selected 50--200 stars that are well identified in USNO-B. The residuals 'FORS2'--'USNO-B' were found to have distributions with long non-Gaussian tails, therefore the identification window size was {set to the value that minimised the residuals of }the coordinate systems fit. Considering that the {uncertainties} of {\small FORS2} positions and proper motions are very small, we temporarily converted our data to the average epoch of each USNO-B star observation, which improved the precision of the coordinate systems alignment. Then we applied the least-squares fit with a model that included full quadratic and cubic polynomials. The typical r.m.s. of individual 'FORS2'--'USNO-B' residuals was $\sim$0.2\arcsec\ and allowed us to transform {\small FORS2} positions to the ICRF system with a precision of  0.05--0.10\arcsec\ for most fields, {  and with an accuracy maintained by 'USNO-B'}. Using the future \emph{Gaia} \citep{Perryman:2001vn} astrometric catalogue, we plan to better tie the {\small FORS2} local fields to the ICRF and to maintain the high precision of $\sim$0.05--0.07~mas for stars of $I=15$--17 magnitude in an absolute sense.

\subsection{Pixel scale}{\label{pix}}
An accurate pixel scale value is necessary to convert astrometric parameters from pixel to the arc metric with no loss of accuracy. The approximate value of 0.126\arcsec px$^{-1}$ given in the {\small FORS2} documentation is not sufficient, as we need to know it with a relative precision of at least $10^{-3}$ $\simeq$ 0.0001\arcsec px$^{-1}$ to match the precision of the parallax measurements. At this level of precision, it should also be known { individually for each sky field, because the fields have reference frames composed in unique ways, they were observed in different temperature conditions with varying fine alignment of the telescope optics, and they are based on approximate star positions.}

The pixel-scale values listed in Table\,\ref{scale} were determined from the derivatives of the transformation fit function along the coordinate axes at the target's position, with a subsequent average of these two values. The scale variation across target fields is slightly larger then the uncertainties and possibly displays a real variation between the reference frames formed in different ways. To
{ convert the astrometric parameters} we applied individual pixel scale values for each field.

{ We ascertained also that the median difference between scales for $x$ and $y$ axes is $-0.37 \pm 0.06$~mas/px. Because this is only $3\cdot 10^{-3}$ of the average scale, the corresponding effect for the conversion of astrometric parameters is negligible for the catalogue stars, but it has to be accounted for the UCDs.}

\subsection{Photometric system}{\label{ph}}
Because the observations were not necessarily obtained in photometric conditions, we estimated the star fluxes approximately by computing the sum of counts within the central 11x11 pixel area of the star images, and the magnitude zero-point was determined for each epoch using external catalogues. The best reference photometry is provided by the DENIS catalogue \citep{DENIS}, which contains $I$-band magnitudes obtained with a Gunn-I filter at 820 nm for stars brighter than 18.5~mag. For most fields, this yielded corrections to {\small FORS2} photometry  with a precision of $\pm$0.01--0.02~mag. For fields Nr. 2, 8, and 15, however, none or very few common stars were found. In these cases, we used the average value of the  corrections determined for other fields and  assumed that the zero-point precision is equal to the scatter of individual field corrections ($\pm$0.08~mag).

\begin{figure}[!]
   \centering
\resizebox{\hsize}{!}{\includegraphics*[bb = 54 50 225 243]{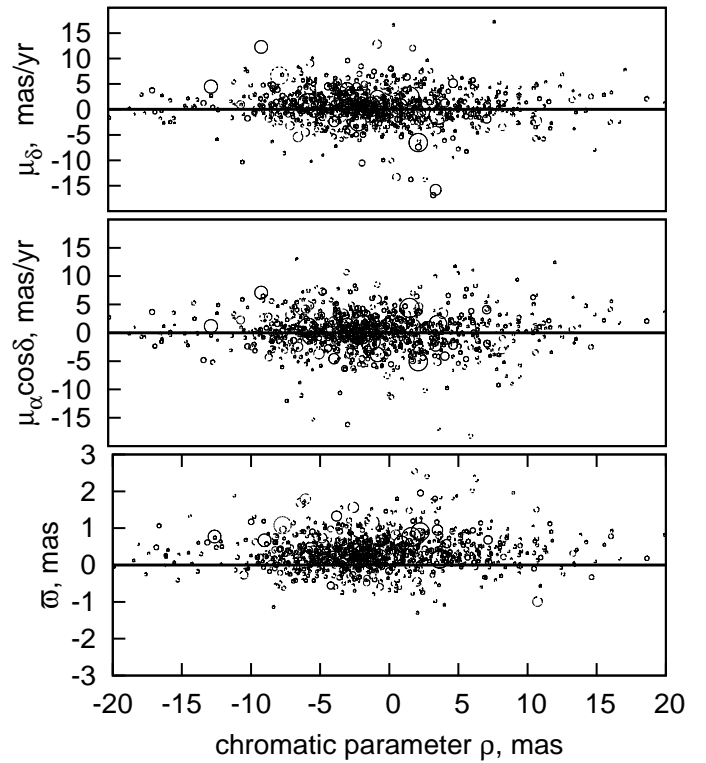}} 
\caption{Distribution of  proper motion and absolute parallaxes on DCR parameter $\rho$ for  bright stars of  the catalogue. Dot size marks the star magnitude.}
\label{ppm_all}
\end{figure}

\subsection{Colour-dependent systematic errors}{\label{dcr_cat}}
{
To control possible colour-dependent systematic errors, we used DCR parameter $\rho$ (Sect.\ref{dcr}), which is +10$\ldots$+20~mas for UCDs and $\rho=0$ corresponds to the average colour of field stars, probably of K--M spectral classes. Parameters $\rho$ are not included in the catalogue because of the large uncertainty of determination, which  is of acceptable $\sim$1--2~mas precision for bright  stars alone.  Using them as equivalent of the colour, we  found  no clear systematic dependence of the proper motion and absolute parallaxes on $\rho$ (Fig.~\ref{ppm_all}) for stars of 15-17.5 magnitude $I$, that is, of the UCDs brightness. 
}

\subsection{Catalogue characteristics}{\label{c_ch}}

The overall catalogue characteristic are given in Table\,\ref{scale} for each target listed with its DENIS identifier: the catalogue star number, number of stars within $\pm 0.5$~mag from the target, which indicates the sky 
density of  bright stars, the size of the reference field $R_{\rm opt}$ for stars in the field centre,  the magnitude limits $I_1$ and $I_2$, the precision of ICRF positions, proper motions, parallaxes, the number of stars in common with 'USNO-B',  the pixel scale, and the r.m.s. of differences FORS2--'USNO-B'.

\begin{table*}[]
\caption [] {Column content of the catalogue  (Sect.\,\ref{c_ch})}
\centering
{\small
\begin{tabular}{@{}cccccccccccccccc@{}}
\hline
\hline
1&2&3&4&5&6&7&8&9&10&11&12&13&14&15&16\\
Nr.& n   &   $I$   &$\sigma_I$& RA   & Dec                 &$\sigma_{\alpha}$&$\sigma_{\delta}$&$\mu_{\alpha}*$ &$\mu_{\delta}$&$\varpi$&$\sigma_{\mu_{\alpha}*}$&$\sigma_{\mu_{\delta}}$&$\sigma_\varpi$&$T$&$\chi^2$\\
  &         & (mag)  &    (mag)      &  (deg)  & (deg)             &(\arcsec)                      &(\arcsec)                   &      (mas                 & (mas)                  & (mas)  & (mas & (mas & (mas) & (yr)  &        \\
  &       &  &       &   &            &                    &                   &      /yr)                 &  /yr)       &   & /yr) & /yr) & &  &   \\
\hline
 1  &   1 & 19.774 & 0.015 &  93.9256115 &  -1.0082455 &   0.148 &   0.116 &     1.67 &    -3.46 &   1.52 &  0.29 & 0.29 & 0.30 & 11.5131 &  1.42  \rule{0pt}{11pt}\\
 1  &   2 & 22.363 & 0.021 &  93.9256675 &  -1.0093558 &   0.146 &   0.115 &    -3.27 &    -0.76 &   2.59 &  1.64 & 1.69 & 1.65 & 11.5131 &  2.93 \\                 
 1  &   3 & 17.013 & 0.014 &  93.9263983 &  -1.0111052 &   0.133 &   0.105 &     2.53 &     2.82 &   0.51 &  0.13 & 0.12 & 0.18 & 11.5131 &  3.15 \\                 
 1  &   4 & 21.272 & 0.017 &  93.9264551 &  -1.0027061 &   0.139 &   0.108 &    -2.29 &     1.03 &  -6.25 &  1.06 & 1.09 & 1.05 & 11.5131 &  1.52 \\                 
\hline
\end{tabular}
}
\label{ttttt}
\end{table*}

The catalogue table, available at the CDS, contains the field number (Column 1), the sequential star number $n$ in the field (Column 2), the $I$-band magnitude and its precision $\sigma_I$ (Columns 3,\,4),  RA and Dec for the equinox and epoch J2000.0 (Columns 5, 6), the precisions in RA and Dec (Columns 7,\,8), the relative proper motion $\mu_{\alpha}^*=\mu_{\alpha}\cos \delta$ and $\mu_{\delta}$ per Julian year (Columns 9, 10), the absolute parallax (Column 11), the precisions of these data (Columns 12--14),  the mean epoch $T$ of observations  in Julian years since J2000.0 (Column 15), and the $\chi^2$-value for the epoch residuals to flag the quality of the data (Column 16). An excerpt from the catalogue is given in Table\,\ref{ttttt}.

\begin{figure*}[!]
\resizebox{\hsize}{!}{\includegraphics[bb = 56 174 585 460,clip]{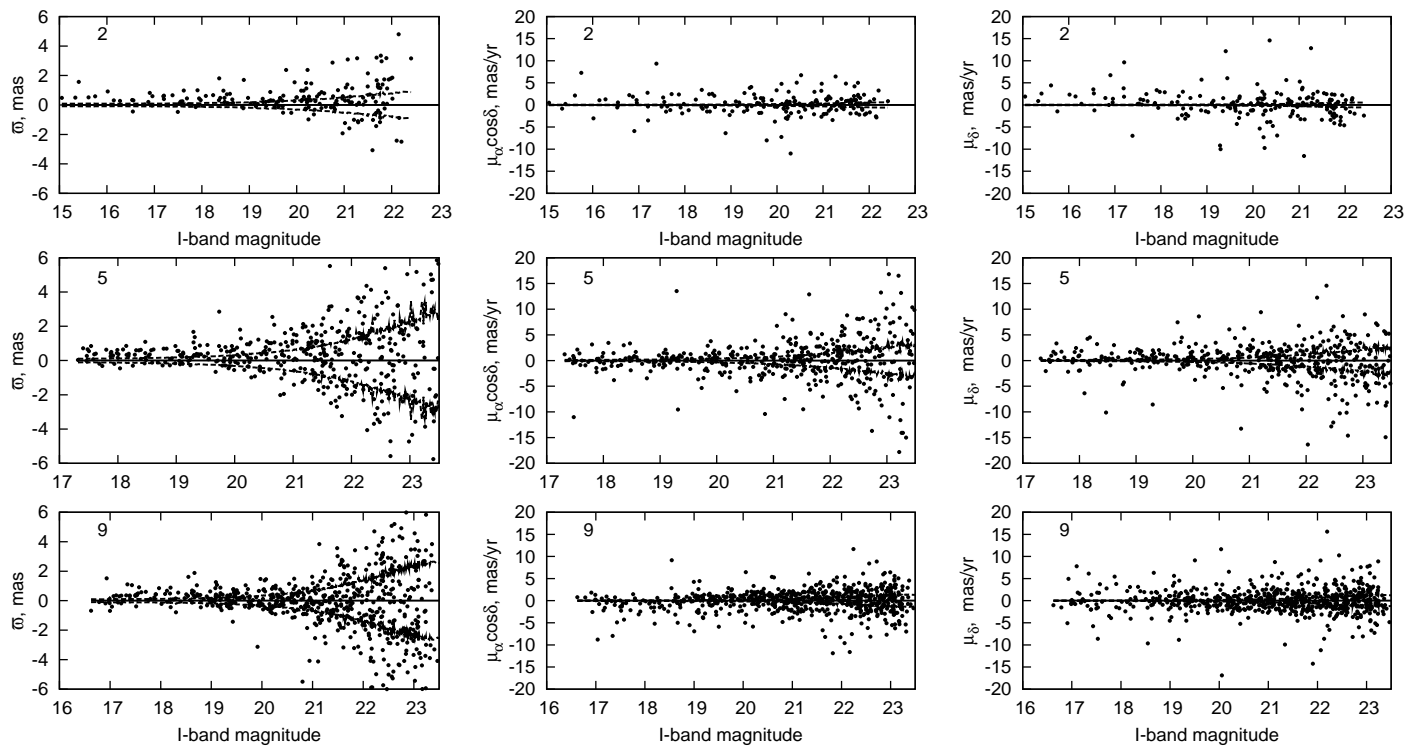}}
\caption {Distribution of absolute parallaxes and relative proper motions of catalogue stars as a function of magnitude in target fields 2, 5, and 9. The sequence number is shown in the upper left corner of the panels and 1-$\sigma$ uncertainty ranges are shown as dashed lines.
}
\label{plt1}
\end{figure*}

The catalogue proper motions are relative and given in the system of distant field stars. As explained in Sect. \ref{field_d}, approximately half of the proper motions are thus negative. The conversion to absolute parallaxes was made using the corrections $\Delta \varpi_{galax}$  derived from the comparison of relative {\small FORS2} parallaxes with a Galaxy model \citep{Robin2003}, which are given in Table~5 of \citetalias{JS2013}.  For bright stars, typical precisions are $\sigma_{\mu}=0.1$~mas/yr and $\sigma_{\varpi}=0.1$~mas  (Table~\ref{scale}), which for stars 5~mag fainter degrades by a factor of ten. For illustration, the distribution of relative proper motions and absolute parallaxes on magnitude is shown in Fig.~\ref{plt1}.
For our example field Nr.\,5 of \dwfive, the parallaxes of relatively bright $I \leq 20$ stars are $\sim 0.5$~mas on average, that is, they are located at about 2~kpc, but some of them with $\varpi=$1--1.5~mas (statistically well over the 3-$\sigma$ margin) are nearby objects at distances of 0.7--1~kpc. From this bright star subsample, one star with $I=19.5$ is located at 250~pc distance. The scatter of proper motions of bright stars is $\pm$ 3--4~mas/yr, which is intrinsic because it significantly exceeds the random {uncertainties}. A few stars have high proper motions of about $\sim$10~mas/yr.

\begin{figure}[!]
   \centering
\resizebox{\hsize}{!}{\includegraphics[bb = 55 49 265 171,clip]{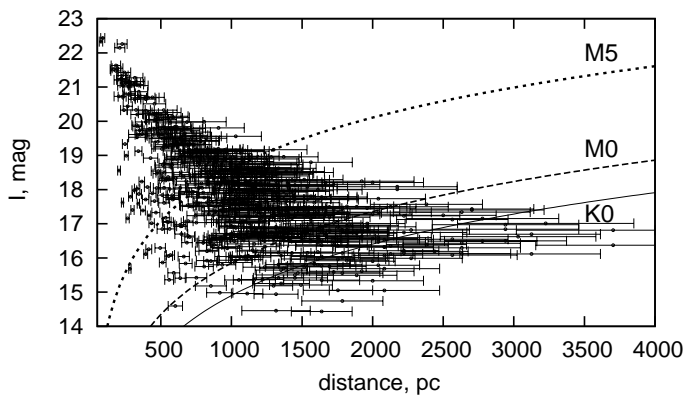}}
\caption {Distribution of distances, with 1-$\sigma$ uncertainty ranges, and $I$-band magnitudes of catalogue stars with well-measured parallaxes.   Theoretical curves correspond to M5, M0, and K0 main-sequence stars. 
}
\label{5pi}
\end{figure}

Like the distribution of $\chi^2$ for the epoch residuals, the distribution of the ratio $\varpi/\sigma_{\varpi}$ has a long non-Gaussian tail. In some instances, parallaxes have negative values much higher than $-3\sigma_{\varpi}$. For bright magnitudes, this is caused by reasons similar to those related to large $\chi^2$ in the epoch residuals (elongated images, etc.) and already discussed in Sect. \ref{conc}. For high star densities (e.g. target fields Nr.17, 18, and 20), where star images overlap frequently, the percentage of large negative parallaxes increases for faint stars. 
{ Fig.~\ref{5pi} shows distances and $I$-band magnitudes of 720 reference stars that have acceptable astrometric solutions ($\chi^2<6$) and well-determined parallaxes ($\varpi/\sigma_{\varpi}>5$). For reference, the theoretical curves corresponding to M5, M0, and K0 main-sequence stars are shown as well. We thus are able to determine reliable distances to $I=15-18$ stars out to a distance of $\sim$1--3.5 kpc.}

\section{Conclusion and discussion}{\label{c_d}}
The {\small FORS2/VLT} observations made for our planet search programme {necessitated} a detailed analysis of the systematic errors in the epoch residuals. We analysed three types of errors: the instability in the relative position of the two CCD chips, systematic errors correlated in space, 
and errors not correlated in space. The chip motion instability and space-dependent errors are relatively well modelled and suppressed to $\sim$0.05~mas. The residual error $\varphi$ of uncertain origin, however, could not be removed. We  only found its r.m.s. value and added it quadratically to the model precision. This component largely determines the single-epoch precision for bright field and dwarf stars, which  in the field centre is 0.12--0.17~mas with a median of 0.15~mas. The median r.m.s. of the epoch residuals is a factor $\theta$ smaller because of the noise absorption by the least squares fitting and is 0.126~mas (0.10--0.15~mas for different fields). These estimates refer to the median seeing of 0.6\arcsec and 32 exposures in one epoch.

\begin{figure}[htb]
   \centering
\resizebox{\hsize}{!}{\includegraphics*[bb = 56 61 261 160]{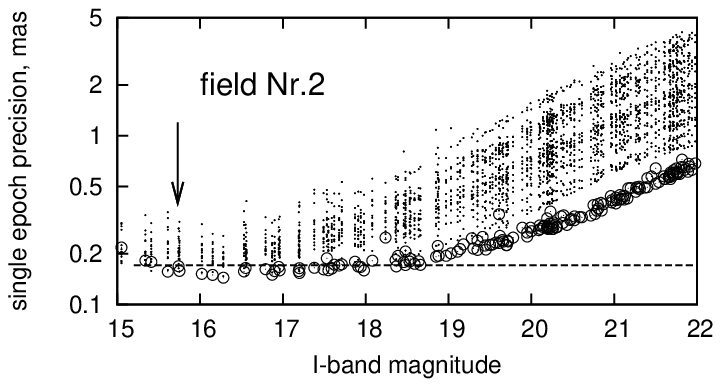}} \\
\resizebox{\hsize}{!}{\includegraphics*[bb = 56 61 261 160]{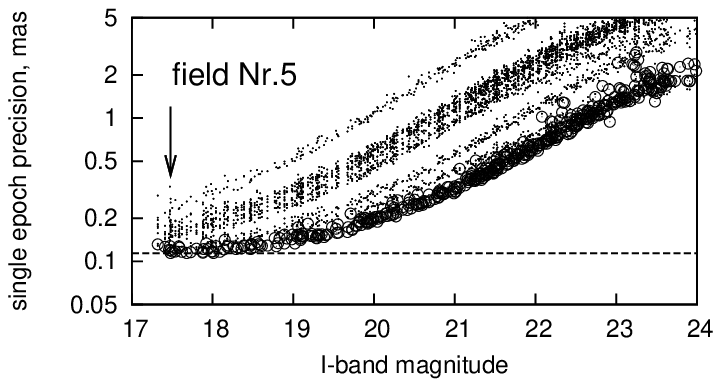}} \\
\resizebox{\hsize}{!}{\includegraphics*[bb = 56 51 261 160]{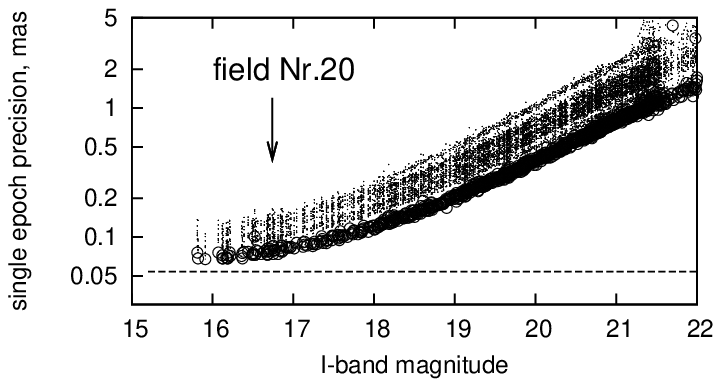}} \\
\caption {Nominal single-epoch precision  $\sigma_{e}/{\theta_{e}}$ as a function of $I$-band magnitude for catalogue stars in fields Nr. 2, 5,  and 20 at any seeing (dots) and for  best-seeing epoch (open circles). The target magnitude is marked by an arrow and the precision floor set by systematic errors $\langle \phi_e \rangle$ is shown with a dashed line. }
\label{gaia}
\end{figure}

The single-epoch precision for bright  stars depends on the field and varies by a factor of two from the best (densely populated but not crowded field Nr.\,20) to typical (Nr.\,5, which is \dwfive), and the worst-case (relatively poor populated star field Nr.\,2) precision. This is shown in Fig. \ref{gaia}, which presents the nominal precision $\sigma_{e}/{\theta_{e}}$ corresponding to an infinite number of epochs. The {vertical}-strip structure of the data points in the plot (each strip corresponds to a separate epoch) is due to the difference in epoch seeing, which varies from 0.4 to 0.9\arcsec. The epochs with best seeing of about 0.40--0.45\arcsec are shown separately by open circles for each field where they form the lower bound with the best precision. In these cases, the nominal {\small FORS2} precision is 0.07--0.15~mas at $I=$16--17~mag.  This demonstrates that the nominal epoch precision improves almost proportionally with seeing, and good seeing is favourable for astrometry.  In these epochs, however, bright reference star images were often saturated and, even with a single saturated pixel, rejected (Sect. \ref{obs}). The number of suitable images within these epochs decreases, and the epoch precision degrades. Moreover, the rejection of equally bright nearby reference stars increases the reference star noise, which deteriorates the precision even more. The result is an upward trend in precision, most distinct for stars brighter than $I=$17~mag in field Nr.\,2. For  field Nr.\,20, the images of selected bright stars never saturate because of stable seeing conditions and the precision improves with brightness. 

The precision floor due to systematic errors is set by  $\phi^2_e$, which is the squared sum of $\varphi$ and the space-correlated error $\phi_{\rm space}(e)$  (Eq. (\ref{eq:upd})). After averaging over all epochs and stars, we obtained the limiting precision $\langle \phi_e \rangle$  marked by the horizontal dashed line in Fig. \ref{gaia}. However, for stars with small space-correlated errors, the precision is better, and the same is true when the small-scale corrections are not measurable due to insufficient number of nearby reference stars and therefore were set to zero.

The current  0.15~mas estimate of the nominal epoch precision for bright field stars and UCDs specifies the astrometric precision of  {\small VLT/FORS2}  routine observations under variable observing conditions, with a restriction of 0.9\arcsec\ on seeing alone. Not surprisingly, this precision exceeds the 0.05~mas value obtained in \citetalias{Lazorenko2009} for observations in optimal conditions.  Here, because of problems with saturation (Sect. \ref{satur}), we restricted the exposure duration, and therefore the light flux for a single image of a bright star is 0.5--0.7$\times 10^6$ photoelectrons, while the peak flux in the pilot study in \citetalias{Lazorenko2009} was 1.5--2.0$\times 10^6$  photoelectrons. Thus, the label 'bright star' really refers to objects with threefold difference in brightness between both studies. For this reason, the median   uncertainty  of the photocentre determination increases from 0.23~mas in \citetalias{Lazorenko2009} to 0.49~mas in the current programme, and the reference frame noise increases respectively from 0.15~mas to 0.30~mas. With this two-fold difference, the pilot study estimates scaled to the current observing conditions leads to 0.10~mas epoch precision, in agreement with Fig.\,16 in \citetalias{Lazorenko2009}, and which is similar to the 0.13~mas precision  for observations  restricted to the optimal seeing (Sect.\ref{av_conc}).

We compared the astrometric performance of the VLT and the \emph{Gaia} predictions, which illuminates effects related to differences in the entrance apertures and consequently in the light collecting power. Astrometry of bright objects is surely much better from dedicated space satellites, but the precision for very faint stars favours the use of large ground-based facilities. We illustrated this by comparing the \emph{Gaia} single-epoch precision \citep{GAIA,de-Bruijne:2012kx} with the precision for VLT/{\small FORS2} catalogue stars shown in Fig. \ref{gaia}, given by {the estimator $\sigma_{e}/{\theta_{e}}$ that does not depend on the number of epochs}. For bright stars, \emph{Gaia} can achieve $\sim$0.01~mas single-epoch precision, while at the faint end of $G=20$, this value is expected to be $\sim$0.7--1.1~mas. With the relation $G$$-$$I$ $\sim$ 0.8$-$1.8, valid for spectral classes F8\ldots L2 \citep{Jordi2010, Jordi2010yCat}, this limit corresponds to the range $I=18.2-19.2$ where {\small FORS2} provides us with an epoch precision of 0.2--0.3~mas. For UCDs with $I$-band magnitudes of 16--18, the {\small FORS2} precision is thus up to five times better than the expected \emph{Gaia} precision. This can be reformulated differently: for faint stars, the expected \emph{Gaia} single-measurement precision can be reached with {\small FORS2} for stars that are approximately 4 magnitudes fainter.

The main motivations of this study was to provide data with well-predicted statistics, which is critically needed for the detection of low-mass companions to UCDs. We {achieved this goal}, because the obtained epoch residuals are distributed according to a normal law with a scatter parameter equal to the model value $\sigma_{e}$. In addition, the $\chi^2$-values for the epoch residuals corrected with the factor $ c_{\chi^2}^2(I) $ follow the theoretical $\chi^2$-statistic.

{ The immediate scientific result of {\small FORS2} high precision astrometry is 
the discovery of two tight binaries, which is discussed in \citetalias{JS2013}, and the 
measurements of trigonometric parallaxes of 20 UCDs and hundreds of $I =$16-17.5 
stars with precision of $\sim 0.1$~mas. This is an excellent performance for 
ground-based optical astrometry and comes close to the best precisions 
obtained in the optical with {HST/WFC3}-scanning \citep{HST} and for 
radio sources with VLBI \citep{VLBI}.}

\begin{appendix}{\label{A}}

\section{Small-scale correlation in epoch residuals}{\label{B2}}
Let $x_e$ and $ x_{e,\ast} $ be the {residual} values {for the epochs $e=1 \ldots N_e$ } introduced in Section \ref{cor}. These values are correlated due to space-dependent errors, thus they contain some common component $\alpha_e$. Formally, this is expressed by 
\begin{equation}
\begin{array}{ll}                               
\label{eq:apB_1}
   x_e &=         \hat x_e           + \alpha_e  \\
   x_{e,\ast}&= \hat x_{e,\ast}   + \alpha_e, 
\end{array}
\end{equation}
where $\hat x_e$ and $\hat x_{e,\ast}$ are uncorrelated random values with zero expectation and with variances equal to the model epoch variances $\sigma_e^2$ and $\sigma^2_{e, {\ast}}$, respectively. We also assume that $\hat x_e$, $\hat x_{e,\ast}$, and  $\alpha_e$ are not correlated. The model (\ref{eq:apB_1}) {implies} that the measured variances
\begin{equation}
\begin{array}{ll}                               
\label{eq:apB_1b}
   D_e&=         \sigma_e^2           +  \alpha_e^2  \\
   D_{e,\ast}&=\sigma^2_{e, {\ast}}    + \alpha_e^2
\end{array}
\end{equation}
for $x_e$ and $x_{e,\ast}$, respectively, are larger than the model predictions. We want to combine $x_e$ and $ x_{e,\ast}$ in a term
\begin{equation}
\label{eq:apB_2}
   A_e=  x_e + a x_{e,\ast}  
\end{equation}
that has minimum variance $\sigma^2(A_e)=E[A_e^2]$ reached for some $a$ (here $E$ denotes mathematical expectation substituted by the average over epochs). This is equivalent to the condition $ \partial E[A_e^2] / \partial a=0$,  which has the solution
\begin{equation}
\label{eq:apB_3}
   a =  - c^2 / (\sigma^2_{e, {\ast}} +c^2),
\end{equation}
yielding the best variance of $A_e$
\begin{equation}
\label{eq:apB_4}
   \sigma^2(A_e) =  D_e - a^2 / D_{e,\ast}
\end{equation}
or
\begin{equation}
\label{eq:apB_5}
   \sigma^2(A_e) =  \sigma_e^2 + 
 a^2  \frac{ \sigma^2_{e, {\ast}} }{ \sigma^2_{e, {\ast}} +c^2 }.
\end{equation}
Here, $c^2=E[x_e x_{e,\ast}]$ corresponds to the covariance between $x_{e}$ and $x_{e,\ast}$ in the calibration file data. Note that $c^2$ and the solution $a$ in Eq.~(\ref{eq:apB_3}) depend on  $r_{\rm small}$ (the radial size containing the calibration stars), therefore $\sigma^2(A_e)$ {has the lowest value for a $r_{\rm small}$ that ensures maximum }$a^2$ or $c^2$. Thus, the optimal $r_{\rm small}$ is defined by the condition on the highest measured correlation $\hat \rho= \langle x_e x_{e,\ast}/( \sigma_e \sigma_{e,{ \ast}}) \rangle $, where the average is taken over epochs. For this specific value of $r_{\rm small}$, and using the measured value of $\hat \rho$, we recover the variance
\begin{equation}
\label{eq:apB_alf}
   \alpha_e^2  = \hat \rho   \sigma_e \sigma_{e,{ \ast}} 
\end{equation}
of the systematic component between epochs.

The relation $\sigma^2(A_e) < D_e$, which follows from Eq.~(\ref{eq:apB_4}), means that the correction never degrades the precision of the result, even for very large $\sigma_{e,{ \ast}}$.  For $\sigma_{e,{ \ast}}\to \infty $, no improvement is expected, because $a \to 0$.

We conclude that the best correction for the {small}-scale space errors to the measured values of $x_e$ is  {not $\alpha_e$ but}
\begin{equation}
\label{eq:delta}
  \Delta'_e=a x_{e,\ast},
\end{equation}
which is determined with the precision $\sigma(A_e)$ (in Sect. \ref{cor} denoted as $\phi_{\rm small}(e)$). It is important that the correction Eq. (\ref{eq:delta}) efficiently mitigates the common systematic component $\alpha_e$. Taking into consideration Eq. (\ref{eq:apB_1}), we find that the initial expectation of $x_e$ is $E[x_e]=\alpha_e$. Following Eq. (\ref{eq:apB_2}), we find that after the correction
\begin{equation}
\label{eq:apB_6b}
  E[A_e]=\alpha_e (1+a).
\end{equation}
The systematic error thus {vanishes} either in the trivial case of $\alpha_e =0$  (no error) or if $a=-1$. The complete subtraction of systematic errors (the case of $a=-1$) is never possible, because $\sigma^2_{e, {\ast}}>0$.
For stars in the field centre, $a$ is usually $-0.2 \ldots -0.7$. For \dwfive, {for example}, $a \approx -0.6$, which means that the systematic errors in this case are reduced by half.

The computation of $ \Delta'_e$ should be based on the calibration files related for the current mode $k$, and ideally, these files should be generated for each $k$. However, this is time consuming and, in practice, we computed the calibration file with $k=10$ alone. In addition, instead of adding $ \Delta'_e$ to $x_e$, whose value depend on $k$, we added it to $X_m$ for each $m \in e$. In the limit $N_e \to \infty$, {both procedures are equivalent}. We verified numerically that with the limited number of epochs given in {our project, the differences between the two procedures are negligible}.

\section{Correlation of individual frame measurements}{\label{B3}}
Because $E[x_e]=\alpha_e $ (see Appendix \ref{B2}), each frame measurement is biased by this constant value, thus $E[x_m]=\alpha_e$. The systematic errors therefore induce the covariance $E[x_m,x_m']=\alpha_e^2$ between the frames $m,m' \in e$. After the correction Eq. (\ref{eq:apB_2}), the covariance between frame measurements within the epoch $e$ is decreased to ${\rm cov}(m,m')=E[(x_m+a x_{e,\ast} )(x_{m'}+a x_{e,\ast} )] = \alpha_e^2 (1+a)^2$, hence $ {\rm cov}(m,m')=  \phi^2_{\rm small}(e)$, due to {small}-scale systematic errors. The effect of large-scale systematic errors is similar, because they are removed incompletely.  The uncertainty $\phi_{\rm long}(e)$ of that correction gives the measure of the bias in the frame measurements, which results in the covariance $ {\rm cov}(m,m')=  \phi_{\rm long}(e)^2$. Taking into account the contribution from $ \varphi$, the complete covariance is
\begin{equation}
\label{eq:apB_6}
  {\rm cov}(m,m')=  \phi^2_e.
\end{equation}
The covariance matrix  $\vec {D}$  (Sect.\ref{layout}) is therefore non-diagonal and has a simple block structure with equal non-diagonal elements $D_{m,m'}= \phi_e^2$ for $m,m' \in e$ and $D_{m,m'}=0$  for $m,m' \not \in e$. The diagonal elements are $D_{m,m}= \sigma_m^2$. For instance, for a two-epoch data set with three frames each, the covariance matrix would be

$\vec {D} = 
\begin{pmatrix}
\sigma_1^2  &  \phi_1^2   &  \phi_1^2  &    0  &  0  &  0  \\
\phi_1^2    &  \sigma_2^2 & \phi_1^2   &    0  &  0  &  0  \\
 \phi_1^2  &  \phi_1^2   & \sigma_3^2 &    0  &  0  &  0  \\
    0  &  0  &  0  &      \sigma_4^2  &  \phi_2^2    & \phi_2^2   \\
    0  &  0  &  0  &      \phi_2^2    &  \sigma_5^2  & \phi_2^2   \\
    0  &  0  &  0  &        \phi_2^2  &  \phi_2^2    & \sigma_6^2 \\
\end{pmatrix}.
$

The non-diagonal structure {increases the} computation time when many field stars are computed as targets (each one computed with reference to all other stars using its own matrix $\vec {D}$). It introduces several nested  loops and the computation using the direct inversion of $\vec {D}$ becomes too slow. Therefore we proceed differently when fitting data of the target and reference stars. The target is processed nominally, that is, using the generalised least-squares fit and the direct computation of the inverse matrix $\vec {D^{-1}}$. For reference stars, 
we used the approximate analytic expression for the inverse matrix
\begin{equation}
\begin{array}{l}                               
\label{eq:t_1}
 D^{-1}_{m,m'}= \frac{-\phi_e^2 \sigma_e''^{-4}}
	{1+\phi_e^2 \sigma_e''^{-2}(n_e-2) -\phi_e^4 \sigma_e''^{-4}(n_e-1) } \\
 D^{-1}_{m,m}=  \sigma_e''^{-2} \frac{1+\phi_e^2 \sigma_e''^{-2}(n_e-2)}
  	{1+\phi_e^2 \sigma_e''^{-2}(n_e-2) -\phi_e^4 \sigma_e''^{-4}(n_e-1) }
  	+ \sigma_m^{-2}  -  \sigma_e''^{-2},	
\end{array}
\end{equation}
where $\sigma_e''^2$ is the weighted average of $\sigma_m^2$ within the epoch $e$. The correcting term  $\sigma_m^{-2}  -  \sigma_e''^{-2}$ added to $ D^{-1}_{m,m}$ partially accounts for the change of precision between the frames. This approximation leads to equal non-diagonal elements in the inverse matrix and increases the computation speed. The expression (\ref{eq:t_1}) is used while running the reduction iterations, whereas the last cycle is performed with the direct inversion of $\vec {D}$.

\end{appendix}

\bibliographystyle{aa}
\bibliography{pa}

\begin{thebibliography}{29}
\expandafter\ifx\csname natexlab\endcsname\relax\def\natexlab#1{#1}\fi

\bibitem[{{Andronov}(1994)}]{Andronov}
{Andronov}, I.~L. 1994, Astronomische Nachrichten, 315, 353

\bibitem[{{Appenzeller} {et~al.}(1998){Appenzeller}, {Fricke}, {F{\"u}rtig},
  {G{\"a}ssler}, {H{\"a}fner}, {Harke}, {Hess}, {Hummel}, {J{\"u}rgens},
  {Kudritzki}, {Mantel}, {Meisl}, {Muschielok}, {Nicklas}, {Rupprecht},
  {Seifert}, {Stahl}, {Szeifert}, \& {Tarantik}}]{FORS}
{Appenzeller}, I., {Fricke}, K., {F{\"u}rtig}, W., {et~al.} 1998, The
  Messenger, 94, 1

\bibitem[{{Avila} {et~al.}(1997){Avila}, {Rupprecht}, \& {Beckers}}]{Avila}
{Avila}, G., {Rupprecht}, G., \& {Beckers}, J.~M. 1997, in Society of
  Photo-Optical Instrumentation Engineers (SPIE) Conference Series, Vol. 2871,
  Society of Photo-Optical Instrumentation Engineers (SPIE) Conference Series,
  ed. A.~L. {Ardeberg}, 1135--1143

\bibitem[{{Benedict} {et~al.}(2010){Benedict}, {McArthur}, {Bean}, {Barnes},
  {Harrison}, {Hatzes}, {Martioli}, \& {Nelan}}]{Benedict:2010ph}
{Benedict}, G.~F., {McArthur}, B.~E., {Bean}, J.~L., {et~al.} 2010, \aj, 139,
  1844

\bibitem[{{de Bruijne}(2012)}]{de-Bruijne:2012kx}
{de Bruijne}, J.~H.~J. 2012, \apss, 341, 31

\bibitem[{{DENIS Consortium}(2005)}]{DENIS}
{DENIS Consortium}. 2005, VizieR Online Data Catalog, 2263, 0

\bibitem[{{ESA}(1997)}]{ESA:1997vn}
{ESA}. 1997, VizieR Online Data Catalog, 1239, 0

\bibitem[{{Jordi} {et~al.}(2010{\natexlab{a}}){Jordi}, {Gebran}, {Carrasco},
  {de Bruijne}, {Voss}, {Fabricius}, {Knude}, {Vallenari}, {Kohley}, \&
  {Mora}}]{Jordi2010}
{Jordi}, C., {Gebran}, M., {Carrasco}, J.~M., {et~al.} 2010{\natexlab{a}},
  \aap, 523, A48

\bibitem[{{Jordi} {et~al.}(2010{\natexlab{b}}){Jordi}, {Gebran}, {Carrasco},
  {de Bruijne}, {Voss}, {Fabricius}, {Knude}, {Vallenari}, {Kohley}, \&
  {Mora}}]{Jordi2010yCat}
{Jordi}, C., {Gebran}, M., {Carrasco}, J.~M., {et~al.} 2010{\natexlab{b}},
  VizieR Online Data Catalog, 352, 39048

\bibitem[{{Lazorenko}(1997)}]{Lazorenko1997}
{Lazorenko}, P.~F. 1997, Kinematics and Physics of Celestial Bodies, 13, 63

\bibitem[{{Lazorenko}(2006)}]{Lazorenko2006}
{Lazorenko}, P.~F. 2006, \aap, 449, 1271

\bibitem[{{Lazorenko} \& {Lazorenko}(2004)}]{LazLaz}
{Lazorenko}, P.~F. \& {Lazorenko}, G.~A. 2004, \aap, 427, 1127

\bibitem[{{Lazorenko} {et~al.}(2009){Lazorenko}, {Mayor}, {Dominik}, {Pepe},
  {Segransan}, \& {Udry}}]{Lazorenko2009}
{Lazorenko}, P.~F., {Mayor}, M., {Dominik}, M., {et~al.} 2009, \aap, 505, 903,
  PL09

\bibitem[{{Lazorenko} {et~al.}(2011){Lazorenko}, {Sahlmann}, {S{\'e}gransan},
  {Figueira}, {Lovis}, {Martin}, {Mayor}, {Pepe}, {Queloz}, {Rodler}, {Santos},
  \& {Udry}}]{Lazorenko2011}
{Lazorenko}, P.~F., {Sahlmann}, J., {S{\'e}gransan}, D., {et~al.} 2011, \aap,
  527, A25

\bibitem[{{Martinez} {et~al.}(2010){Martinez}, {Kolb}, {Sarazin}, \&
  {Tokovinin}}]{seeing}
{Martinez}, P., {Kolb}, J., {Sarazin}, M., \& {Tokovinin}, A. 2010, The
  Messenger, 141, 5

\bibitem[{{Mignard}(2011)}]{GAIA}
{Mignard}, F. 2011, Advances in Space Research, 47, 356

\bibitem[{{Monet} {et~al.}(1992){Monet}, {Dahn}, {Vrba}, {Harris}, {Pier},
  {Luginbuhl}, \& {Ables}}]{Monet1992}
{Monet}, D.~G., {Dahn}, C.~C., {Vrba}, F.~J., {et~al.} 1992, \aj, 103, 638

\bibitem[{{Monet} {et~al.}(2003){Monet}, {Levine}, {Canzian}, {Ables}, {Bird},
  {Dahn}, {Guetter}, {Harris}, {Henden}, {Leggett}, {Levison}, {Luginbuhl},
  {Martini}, {Monet}, {Munn}, {Pier}, {Rhodes}, {Riepe}, {Sell}, {Stone},
  {Vrba}, {Walker}, {Westerhout}, {Brucato}, {Reid}, {Schoening}, {Hartley},
  {Read}, \& {Tritton}}]{USNO}
{Monet}, D.~G., {Levine}, S.~E., {Canzian}, B., {et~al.} 2003, \aj, 125, 984

\bibitem[{{Perryman} {et~al.}(2001){Perryman}, {de Boer}, {Gilmore}, {H{\o}g},
  {Lattanzi}, {Lindegren}, {Luri}, {Mignard}, {Pace}, \& {de
  Zeeuw}}]{Perryman:2001vn}
{Perryman}, M.~A.~C., {de Boer}, K.~S., {Gilmore}, G., {et~al.} 2001, \aap,
  369, 339

\bibitem[{{Pravdo} \& {Shaklan}(1996)}]{Pravdo1996}
{Pravdo}, S.~H. \& {Shaklan}, S.~B. 1996, \apj, 465, 264

\bibitem[{{Pravdo} {et~al.}(2004){Pravdo}, {Shaklan}, {Henry}, \&
  {Benedict}}]{Pravdo2004}
{Pravdo}, S.~H., {Shaklan}, S.~B., {Henry}, T., \& {Benedict}, G.~F. 2004,
  \apj, 617, 1323

\bibitem[{{Reid} \& {Honma}(2013)}]{VLBI}
{Reid}, M.~J. \& {Honma}, M. 2013, ArXiv e-prints

\bibitem[{{Riess} {et~al.}(2014){Riess}, {Casertano}, {Anderson}, {Mackenty},
  \& {Filippenko}}]{HST}
{Riess}, A.~G., {Casertano}, S., {Anderson}, J., {Mackenty}, J., \&
  {Filippenko}, A.~V. 2014, ArXiv e-prints

\bibitem[{{Robin} {et~al.}(2003){Robin}, {Reyl{\'e}}, {Derri{\`e}re}, \&
  {Picaud}}]{Robin2003}
{Robin}, A.~C., {Reyl{\'e}}, C., {Derri{\`e}re}, S., \& {Picaud}, S. 2003,
  \aap, 409, 523

\bibitem[{{Sahlmann} {et~al.}(2013{\natexlab{a}}){Sahlmann}, {Lazorenko},
  {M{\'e}rand}, {Queloz}, {S{\'e}gransan}, \& {Woillez}}]{Sahlmann_spie}
{Sahlmann}, J., {Lazorenko}, P.~F., {M{\'e}rand}, A., {et~al.}
  2013{\natexlab{a}}, in Society of Photo-Optical Instrumentation Engineers
  (SPIE) Conference Series, Vol. 8864, Society of Photo-Optical Instrumentation
  Engineers (SPIE) Conference Series

\bibitem[{{Sahlmann} {et~al.}(2013{\natexlab{b}}){Sahlmann}, {Lazorenko},
  {S{\'e}gransan}, {Mart{\'{\i}}n}, {Queloz}, {Mayor}, \& {Udry}}]{JS2013}
{Sahlmann}, J., {Lazorenko}, P.~F., {S{\'e}gransan}, D., {et~al.}
  2013{\natexlab{b}}, \aap, 556, A133

\bibitem[{{Sahlmann} {et~al.}(2014){Sahlmann}, {Lazorenko}, {S{\'e}gransan},
  {Mart{\'{\i}}n}, {Queloz}, {Mayor}, \& {Udry}}]{Sahlmann:2013prep}
{Sahlmann}, J., {Lazorenko}, P.~F., {S{\'e}gransan}, D., {et~al.} 2014, \aap,
  submitted, Paper I

\bibitem[{{Seager, S.}(2011)}]{Seager:2011ve}
{Seager, S.}, ed. 2011, {Exoplanets} (University of Arizona Press)

\bibitem[{{Sozzetti}(2005)}]{Sozzetti:2005qy}
{Sozzetti}, A. 2005, \pasp, 117, 1021

\end{thebibliography}

\end{document}